\documentclass[12pt,a4paper]{article}
\usepackage[utf8]{inputenc}
\usepackage{amsmath}
\usepackage{amsfonts}
\usepackage{amssymb}
\usepackage{natbib}
\usepackage[dvips]{graphicx}
\graphicspath{{figures/}}

\title{\bf $\Delta\mu$ Binaries among Stars with Large Proper Motions}

\author{M.Yu. Khovritchev$^{1}$\footnote{e-mail: deimos@gao.spb.ru} \  and A.M. Kulikova$^1$ \\
{\small $^1$Pulkovo Observatory, 65/1 Pulkovskoye chaussee, Saint Petersburg, 196140, Russia}}

\begin{document}

\date{}
\maketitle

\begin{abstract}
Based on observations performed with the Pulkovo normal astrograph in 2008-2015 and data from sky surveys (DSS, 2MASS, SDSS DR12, WISE), we have investigated the motions of 1308 stars with proper motions larger than $300~ mas/yr$ down to magnitude 17. The main idea of our search for binary stars based on this material is reduced to comparing the quasi-mean (POSS2-POSS1; an epoch difference of $\approx$50 yr) and quasi-instantaneous (2MASS, SDSS, WISE, Pulkovo; an epoch difference of $\approx$10 yr) proper motions. If the difference is statistically significant compared to the proper motion errors, then the object may be considered as a $\Delta\mu$-binary candidate. One hundred and twenty one stars from among those included in the observational program satisfy this requirement. Additional confirmations of binarity for a number of stars have been obtained by comparing the calculated proper motions with the data from several programs of stellar trigonometric parallax determinations and by analyzing the asymmetry of stellar images on sky-survey CCD frames. Analysis of the highly accurate SDSS photometric data for four stars (J0656+3827, J0838+3940, J1229+5332, J2330+4639) allows us to reach a conclusion about the probability that these $\Delta\mu$ binaries are white dwarf + M dwarf pairs.
\\\
\end{abstract}


\section*{Introduction}

The investigation of stellar populations within the nearest 50 pc of the Sun arouses great interest in the astronomical community. This is largely because a detailed study of the kinematics and statistical properties of a stellar ensemble in this region of the Galaxy makes it possible to gain an insight into the genesis of the circumsolar region and to test the models of Galactic structure and evolution. This is reflected in a whole series of papers (see, e.g., \citet{Zenoviene2015}; \citet{Sanders2015}). Dwarf stars (predominantly M dwarfs, subdwarfs, white and brown dwarfs), which constitute the main population of the solar neighborhood, are noticeably brighter than more distant objects of this type, which makes them the most convenient and often the only accessible targets for studies.

Such empirical relations as mass–luminosity and mass–radius are important tests of the models for the structure of stars and their atmospheres. Whereas theoretical models for comparatively massive stars (1–3 $M_\odot$ ) show good agreement with observations, there are difficulties for low-mass stars (see, e.g., \citet{Spada2013}). They can be overcome in part by improving the mentioned relations. This requires revealing a large number of binary systems and determining their orbits.

The parameters of the mass function and the fraction of binary system in a given stellar population can be estimated on the basis of cosmogonic models. Comparison of such estimates with observations enables a justified choice of the most probable scenario for the formation and evolution of stellar populations to be made (\citet{Thies2015}). Therefore, revealing binary systems among low-luminosity objects is an important research direction.

The topicality of the problems under consideration contributed to the implementation of several projects in the 20th century aimed at detecting fast stars (\citet{Dunn1995}; \citet{Luyten1979}) and determining their proper motions and trigonometric parallaxes (\citet{Altena1995}). As a rule, the stars in the solar neighborhood are distinguished by large proper motions 
($\mu>$ 0.1 arcsec$\cdot$yr$^{-1}$ ). 
This has allowed a sample of low-luminosity objects to be produced by detecting such relatively fast objects on the basis of astrometric observations. Not by chance, the LSPM catalog of fast stars (\citet{Lepine2005}) is invoked for the selection of low-luminosity objects more often than others.

The passage to CCD detectors has rekindled interest in this subject matter. At present, several research groups in the world are actively working in this direction. For example, the RECONS~\footnote{http://www.recons.org} group (Riedel et al. 2014) or the MEarth project (\citet{Dittmann2014}). A project to investigate stars with large proper motions was also implemented at the Pulkovo Observatory (\citet{Khovritchev2013}).

The GAIA space observatory has been operating at the Lagrange point L2 already for more than a year. The main product of this space mission is an astrometric catalog (coordinates, velocities, distances, photometric data) of a billion stars down to 20$^{m}$. The final accuracy of the astrometric parameters will be unprecedentedly high (for single 16 m stars, $\sim$30 $\mu$as for the positions and 25 $\mu$as$\cdot$yr$^{-1}$ for the proper motions). This will allow an enormous number of binary systems containing low-mass components to be detected and investigated (\citet{Soderhjelm2005}). In many cases, reliable orbits will be constructed and the masses will be determined.

Note that the revolution periods for a large number of binary systems are considerably longer than the operation time of the GAIA space telescope. Therefore, a combination of ground-based and space observations performed at different epochs and GAIA data seems promising for a global search of astrometric binaries among low-luminosity stars (\citet{Michalik2014}). Such a combination of data from the FK5 and Hipparcos catalogs has already shown the efficiency of this approach for analyzing the motions of the brightest stars in the sky (\citet{Wielen1999}).

Traditionally, to investigate the motions of the components of visual double stars, their observations are regularly carried out over decades (\citet{Izmailov2010}). A careful analysis of the data series often leads to the detection of stars with invisible companions, also known as, astrometric binaries (\citet{Grosheva2006}). For most nearby low-luminosity stars, we have no such dense, long, and homogeneous series of observations. We are dealing with occasional observations from different projects (which makes them inhomogeneous from the standpoint of the accuracy and systematic errors of the stellar coordinates). The idea of searching for astrometric binaries under such conditions was implemented by \citet{Wielen1999} as applied to the brightest stars in the sky (stars from the FK6 catalog). It consists in the fact that the stellar proper motion determined at a large epoch difference (50–100 yr) reflects the mean motion of the photocenter of an unresolved stellar pair (quasi-mean proper motion), while the proper motion determined in a time interval of $\sim$10 yr corresponds to the instantaneous motion with some probability (quasi-instantaneous proper motion). A statistically significant difference between these proper motions may suggest that we are dealing with a binary system. Such stars are called $\Delta\mu$-binary candidates in Wielen’s terminology.

Physically, assigning a star to the category of $\Delta\mu$ binaries implies the detection of evidence for nonlinearity of its motion over the celestial sphere. Several factors can be responsible for this effect. A classic example is an optically unresolvable binary system. In this case, we are dealing with a nonlinear motion of the binary’s photocenter relative to its center of mass. A situation where the star being investigated is a component of a wide pair and the second star is too faint to be detectable in the images is possible.

Within the framework of the Pulkovo program of research on stars with large proper motions, we have made an attempt to extend Wielen’s idea to fast stars. To improve the quasi-instantaneous proper motions, we performed astrometric observations of 1123 selected stars from the LSPM catalog. The quasi-mean proper motions were taken from the LSPM catalog. As a result, 70 $\Delta\mu$ binaries were revealed (\citet{Khrutskaya2011}).

A significant shortcoming of the previous study is that the original stellar coordinates from the catalogs and surveys (M2000, CMC14, 2MASS, SDSS DR7) and the LSPM proper motions represent the HCRF/Tycho-2 system only formally. It is natural to expect more reliable results if the proper motions are determined just as was done in photographic astrometry or in the problem of determining the trigonometric parallaxes of stars. In this paper, we attempted to implement precisely this approach, i.e., the CCD frames from different surveys were considered as a set of images entering into a single parallactic series. The parameters of the transition to the reference frame were calculated. The systematic errors of the stellar coordinates were taken into account. Based on the derived set of coordinates for a program star in the system of the reference frame, we calculated the proper motions and quantities that allowed us to reach a conclusion about whether the star belongs to the category of $\Delta\mu$ binaries. Our detailed technique and results are presented in the succeeding sections of this paper.

\section*{Observations and data reduction}
\subsection*{\it Producing the List of Stars and Peculiarities of Our Observations}
As previously (\citet{Khrutskaya2011}), stars from the LSPM catalog were included in our program of observations. The main list contains 1972 stars down to 17$^{m}$ in the declination zone from 30$^{\circ}$  to 70$^{\circ}$. These are mostly stars with $\mu>$ 300 mas$\cdot$yr$^{-1}$ (1507 stars). For various reasons, we added 465 stars. Some of them have large photometric parallaxes; a number of stars were added to analyze the possibility of investigating comparatively slowly moving objects ($\mu>$ 100 mas$\cdot$yr$^{-1}$).

\begin{figure}[h!]
\begin{minipage}[h]{0.5\linewidth}
\center{\includegraphics[width=1\linewidth]{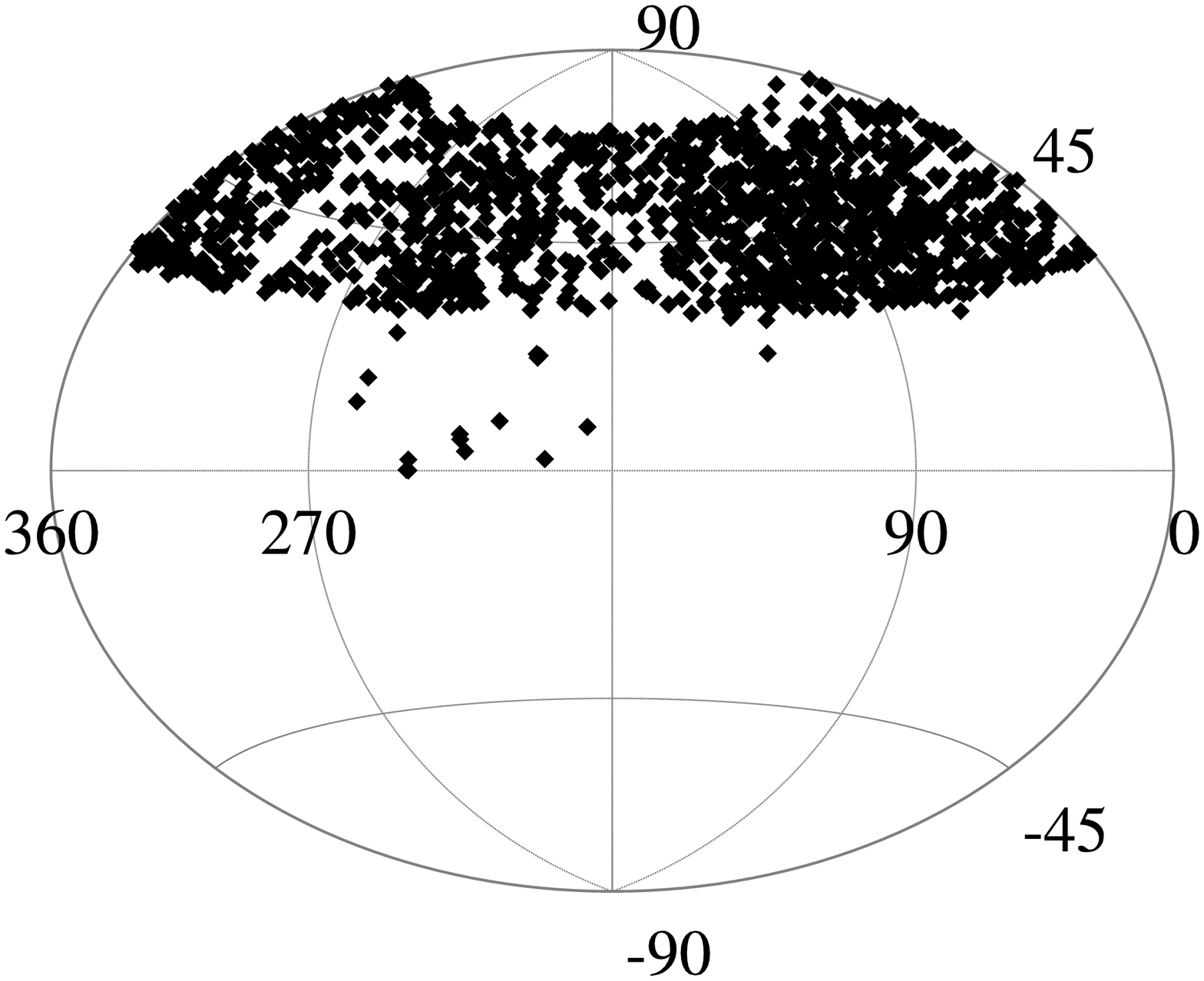} a}
\end{minipage}
\begin{minipage}[h]{0.5\linewidth}
\center{\includegraphics[width=1\linewidth]{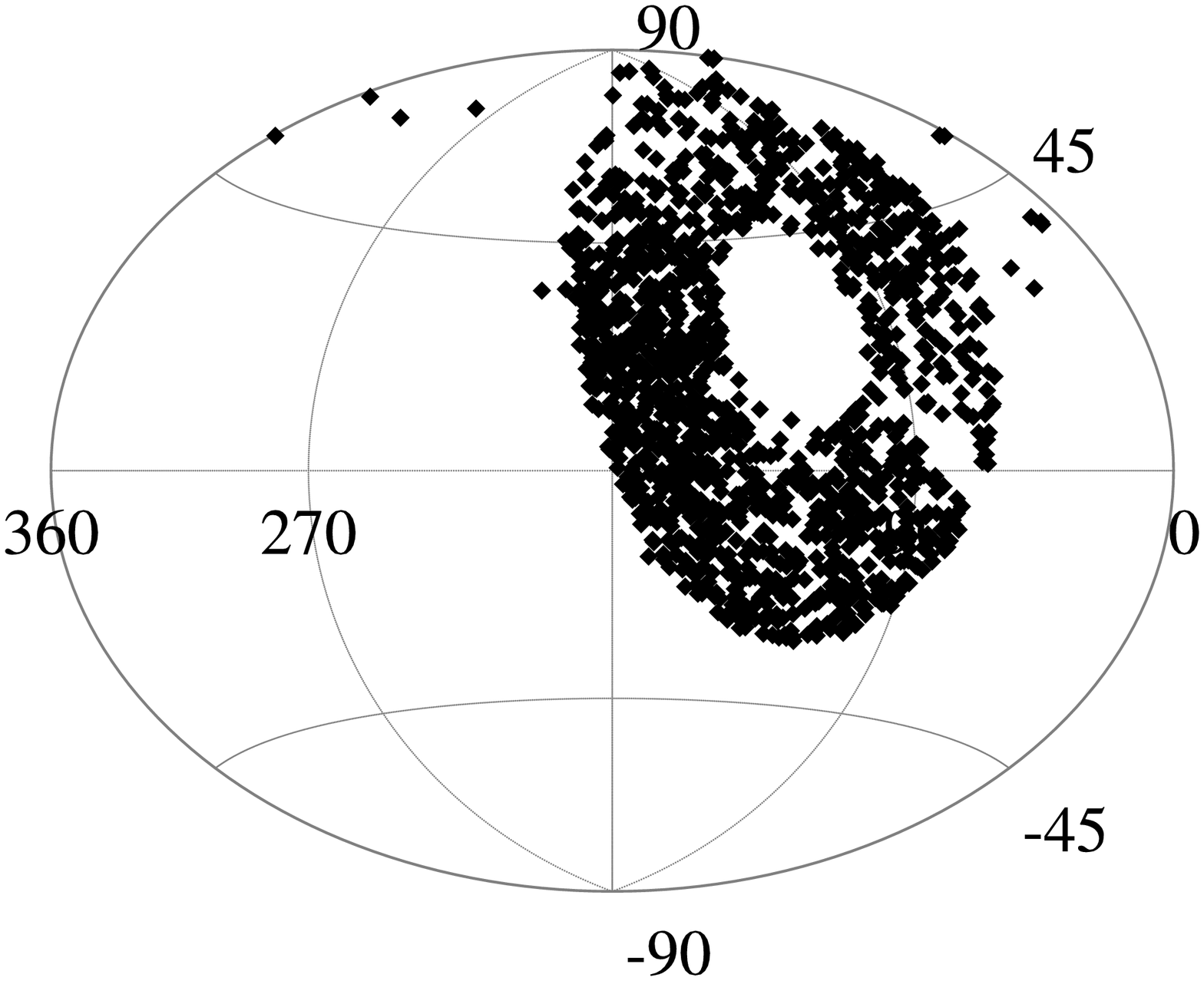} b}
\end{minipage}
\caption{\rm Distribution of Pulkovo program stars over the celestial sphere in the equatorial (a) and Galactic (b) coordinate systems.}
\label{fig1}
\end{figure}

\begin{figure}[h!]
\begin{minipage}[h]{0.5\linewidth}
\center{\includegraphics[width=1\linewidth]{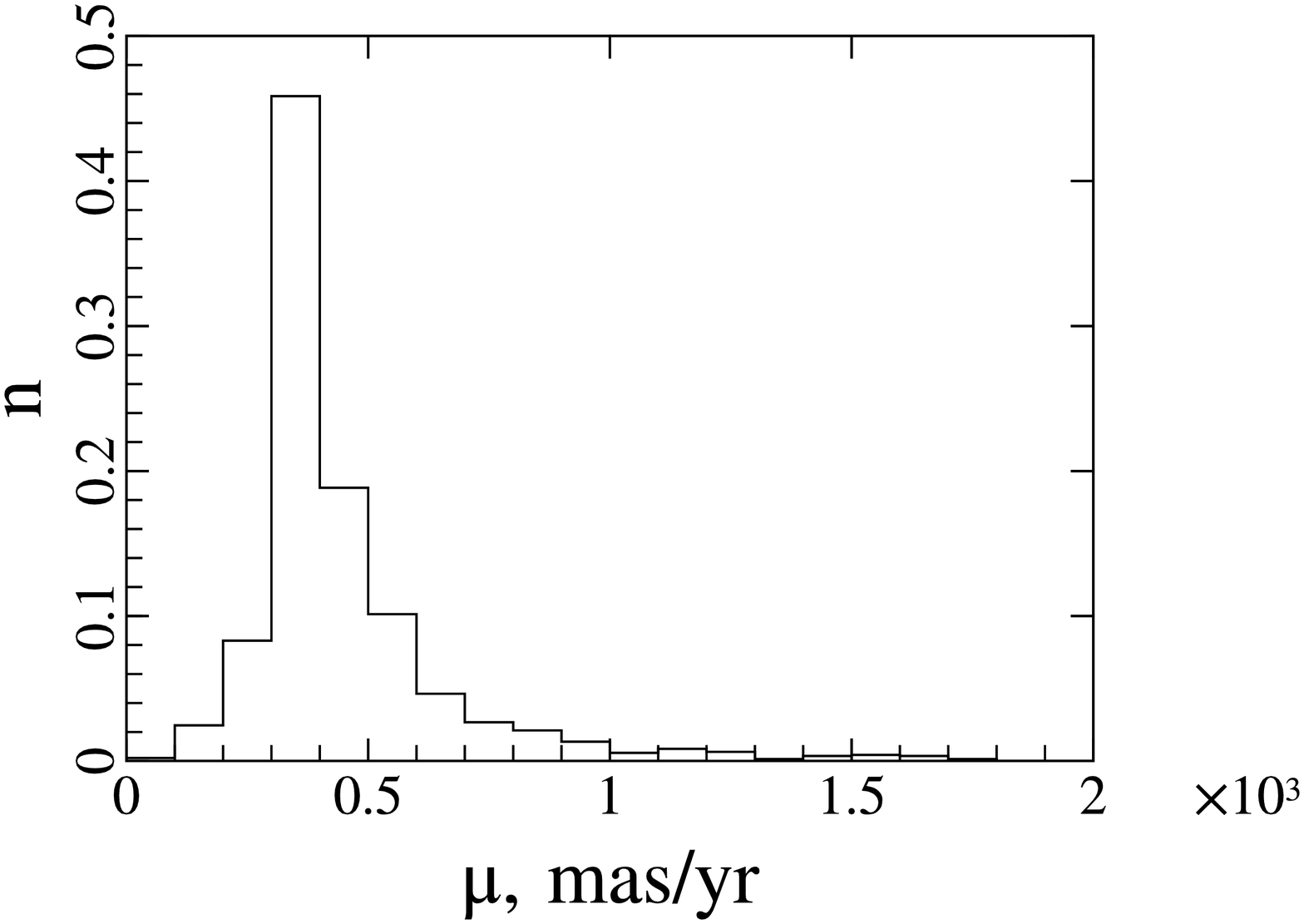} a}
\end{minipage}
\begin{minipage}[h]{0.5\linewidth}
\center{\includegraphics[width=1\linewidth]{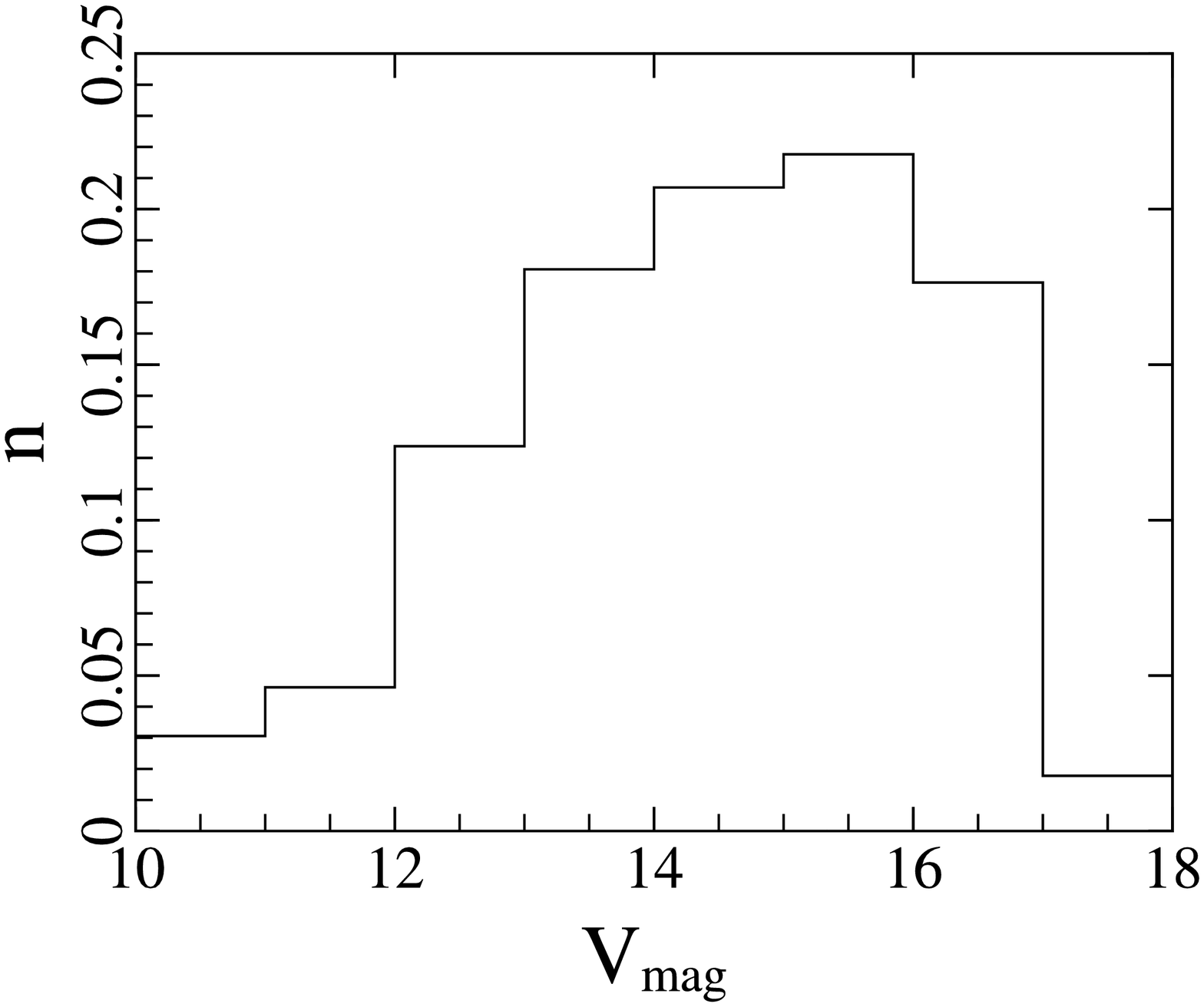} b}
\end{minipage}
\begin{minipage}[h]{0.5\linewidth}
\center{\includegraphics[width=1\linewidth]{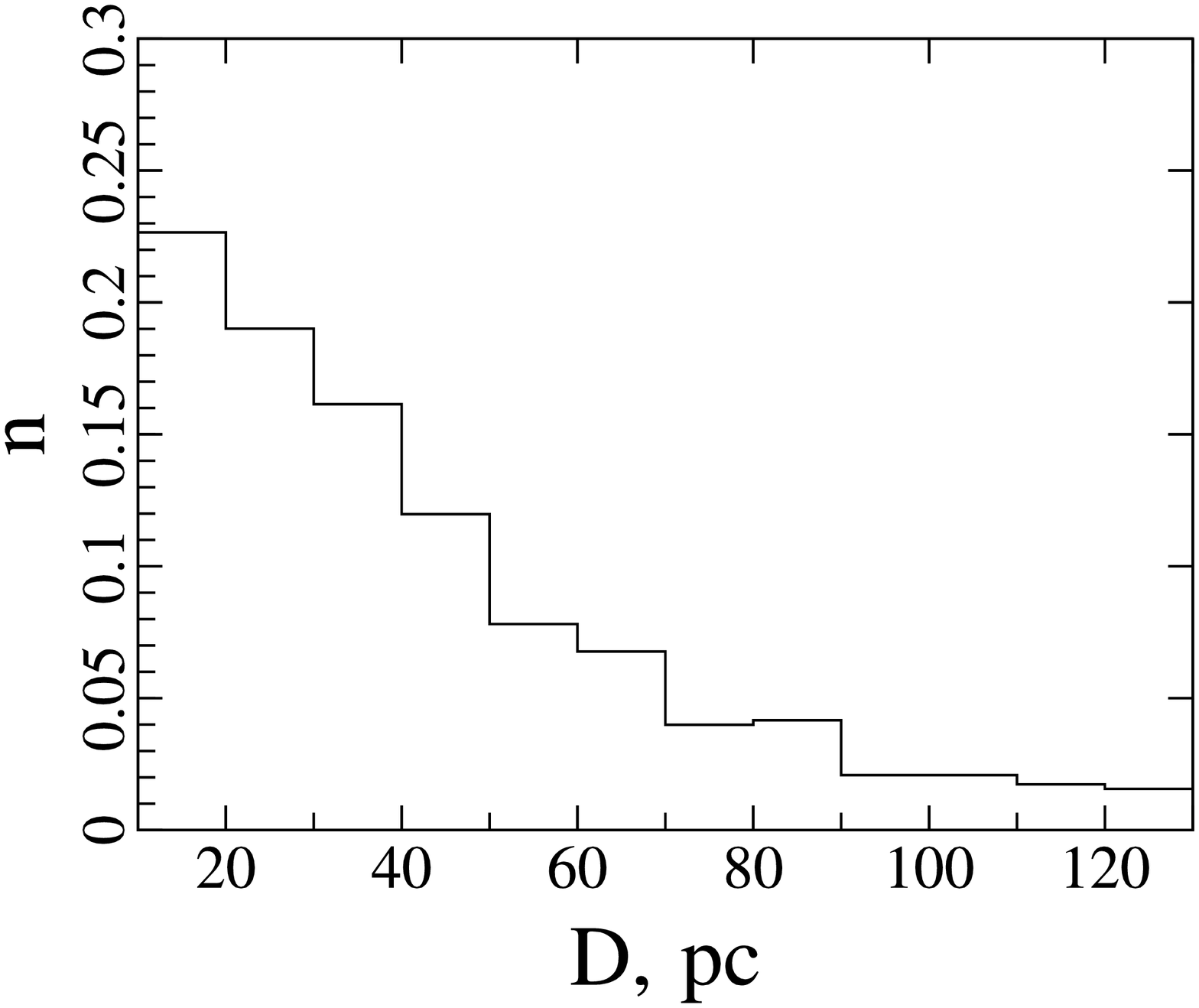} c}
\end{minipage}
\begin{minipage}[h]{0.5\linewidth}
\center{\includegraphics[width=1\linewidth]{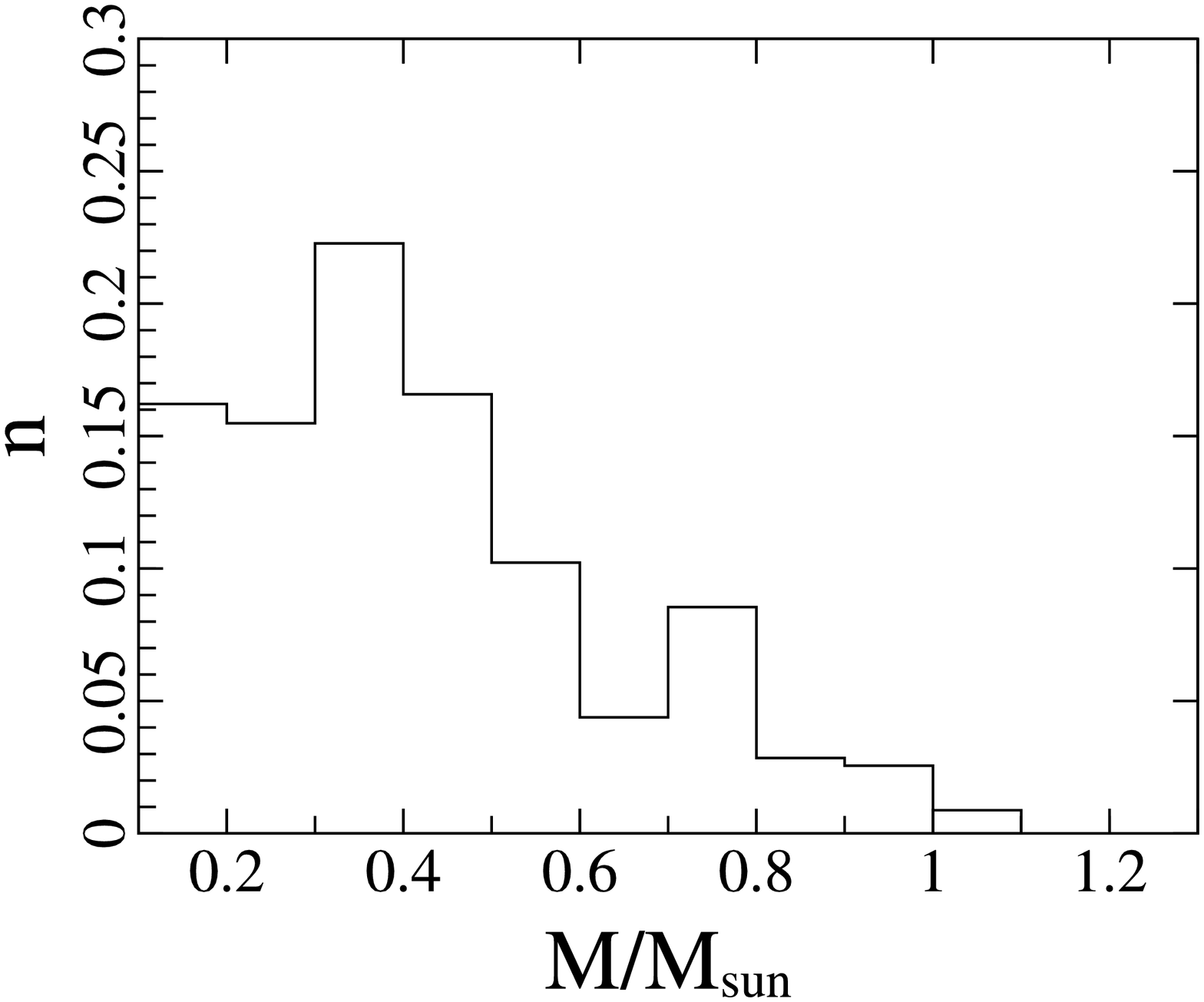} d}
\end{minipage}
\caption{\rm Distribution of Pulkovo program stars in total proper motion (a), magnitude (b), heliocentric distance (c), and mass (d).}
\label{fig2}
\end{figure}

The distribution of observational program stars over the celestial sphere is shown in Fig.~\ref{fig1}. Figure~\ref{fig2} demonstrates how the program objects are distributed in total proper motion, magnitude, mass, and heliocentric distance. On the whole, this is an almost complete sample. The LSPM catalog contains almost all existing fast stars down to $19^m$ (\citet{Lepine2005}). Therefore, it can be said that the Pulkovo observational program includes all of the stars that can be effectively observed in Pulkovo in the corresponding ranges of magnitudes and proper motions.

Our CCD observations were carried out with the Pulkovo normal astrograph (D = 330 mm, F = 3500 mm) from 2008 to 2015. They were necessary to increase the accuracy of the quasi-instantaneous stellar proper motions. An S2C camera (with a working field of $18 \times 16$ arcmin at a scale of 900 mas/pix) was used before 2014. A SBIG ST-L-11K camera (with a working field of $35 \times 23$ arcmin at a scale of 530 mas/pix) was installed in 2014 and 2015. The imaging was performed at hour angles $\pm 1$ h by series from 5 to 10 frames with exposure times of 60 or 120 s, depending on the magnitude of the program star. In several cases, for faint stars the number of frames in a series was increased to 40 to achieve the required signal-to-noise ratio by the summation of individual frames.

\subsection*{\it Using Data from Digital Sky Surveys}

The concept of a virtual observatory, when the results of observations (including the images of sky fields) are accessible via the Internet, has long been used in astronomical practice. The sky surveys containing CCD frames or scans of photographic plates of the entire sky or much of it are of interest to us. In this paper, we used data from SDSS DR12~\footnote{http://dr12.sdss3.org/fields} (\citet{Alam2015}), 2MASS~\footnote{http://irsa.ipac.caltech.edu/ibe} (\citet{Cutri2003}), WISE~\footnote{http://irsa.ipac.caltech.edu/applications/wise} (\citet{Wright2010}), and STScI DSS~\footnote{http://archive.stsci.edu/cgi-bin/dss{\_}form} (\citet{Lasker1998})(scans of POSSI-O and POSSII-J plates).

For all of the listed surveys, there is a convenient interface to request the fits files. The necessary data were downloaded automatically with a specially developed application, and it took several days to do so due to the large volume of WISE data (the fields in this survey were imaged with numerous overlaps). For WISE, we used the frames of only the “cold” part of the mission in the W1 and W2 bands to ensure an acceptable signal-to-noise ratio for the images of faint stars.

\subsection*{\it Determining the Pixel Coordinates}

We analyzed all CCD frames and scans of POSS1 and POSS2 plates according to a unified scheme. A shapelet decomposition (\citet{Massey2005}) was used to fit the stellar images on each frame.

\begin{figure}[h!]
\center{\includegraphics[width=1\linewidth]{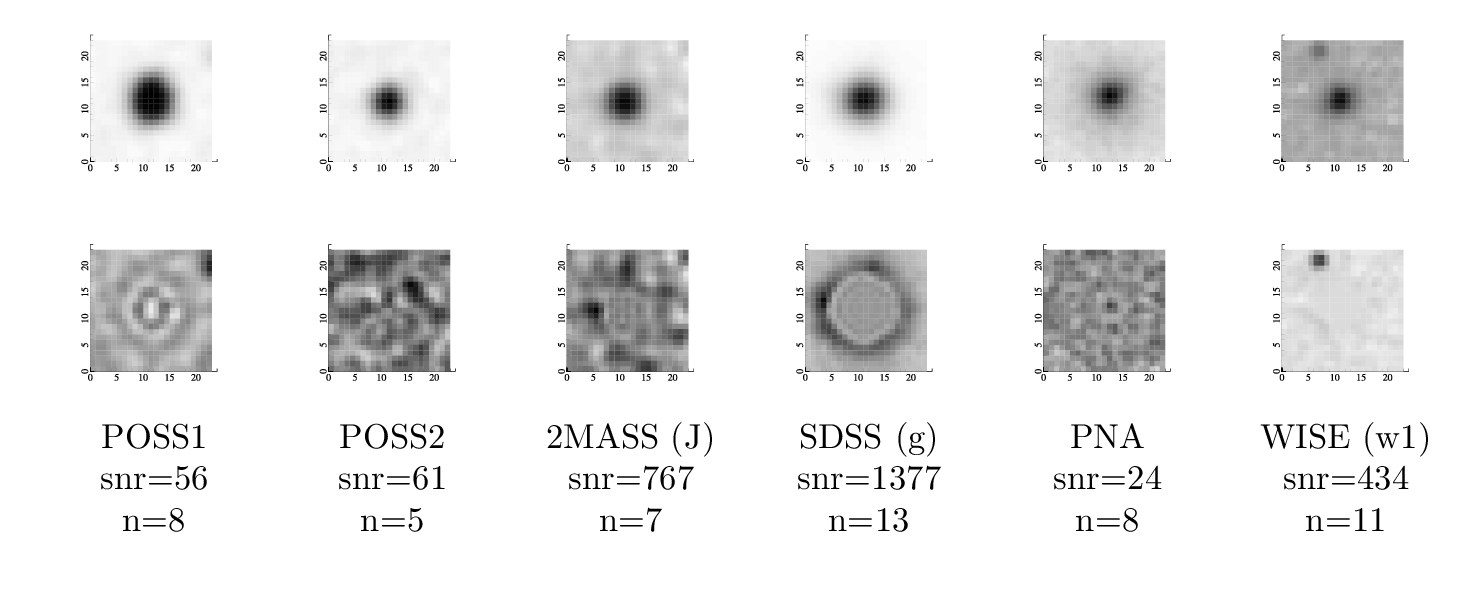}}
\caption{\rm Fitting the image of J0753+5106 ($V_{mag}$=14.2) on the frames from various sky surveys. The upper row contains the original images; the lower row presents the real–model subtraction result. The signal-to-noise ratio for the pixel with maximum intensity (snr) and the shapelet decomposition order (n) are shown. The image region is $24\times24$~pix in size.}
\label{fig3}
\end{figure}

The pixel coordinates of the stellar image photocenter were varied according to the scheme from the paper cited above to ensure the minimum of the sum of the residuals squared. The same principle was applied to automatically select the decomposition order. Typical examples of the fits to the images of a comparatively bright star on the frames from different surveys are shown in Fig.~\ref{fig3}. On the whole, the image measurement technique works reliably. Some problems were observed with the choice of a decomposition order for the SDSS frames. Based on the structure of the distribution of residuals, we can assume that the order was slightly overestimated. As the results of astrometric SDSS frame reductions showed, this is not critical for astrometric measurements (for stars with V$_{mag}$ = 14, the convergence of the pixel coordinates lies within 10 mas).

\subsection*{\it Analysis of the Systematic Errors in the Stellar\\Coordinates from Different Surveys}

The approach relying on the generation of the vector fields of residual pixel stellar coordinate differences based on the entire data set is commonly used to construct astrometric catalogs on the basis of CCD observations. We performed an astrometric reduction for all frames from a given survey. The frames constants allowed us to calculate the residual catalog–frame differences of the pixel coordinates of reference stars.

\begin{figure}[h!]
\begin{minipage}[h]{0.5\linewidth}
\center{\includegraphics[width=1\linewidth]{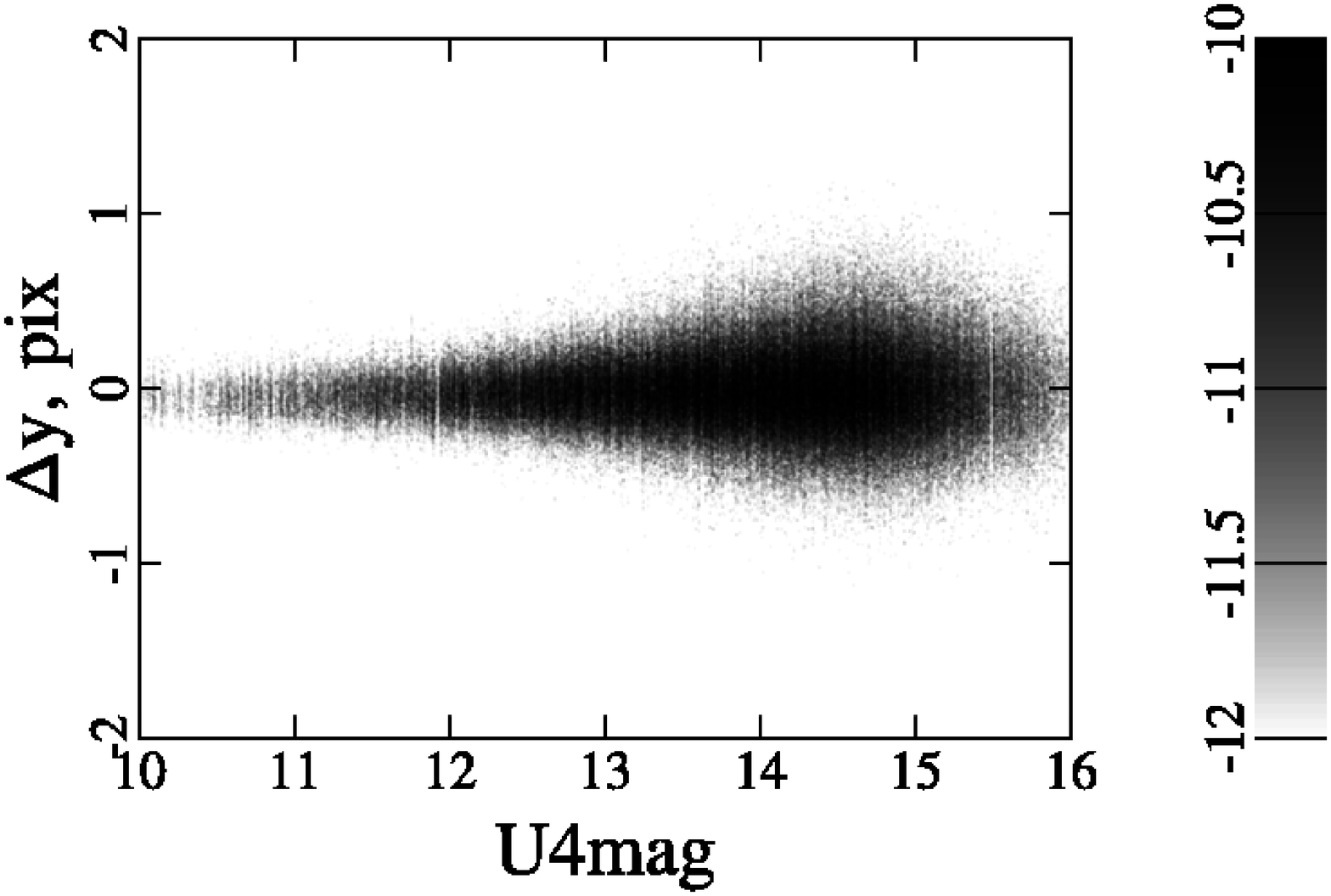} a}
\end{minipage}
\begin{minipage}[h]{0.5\linewidth}
\center{\includegraphics[width=1\linewidth]{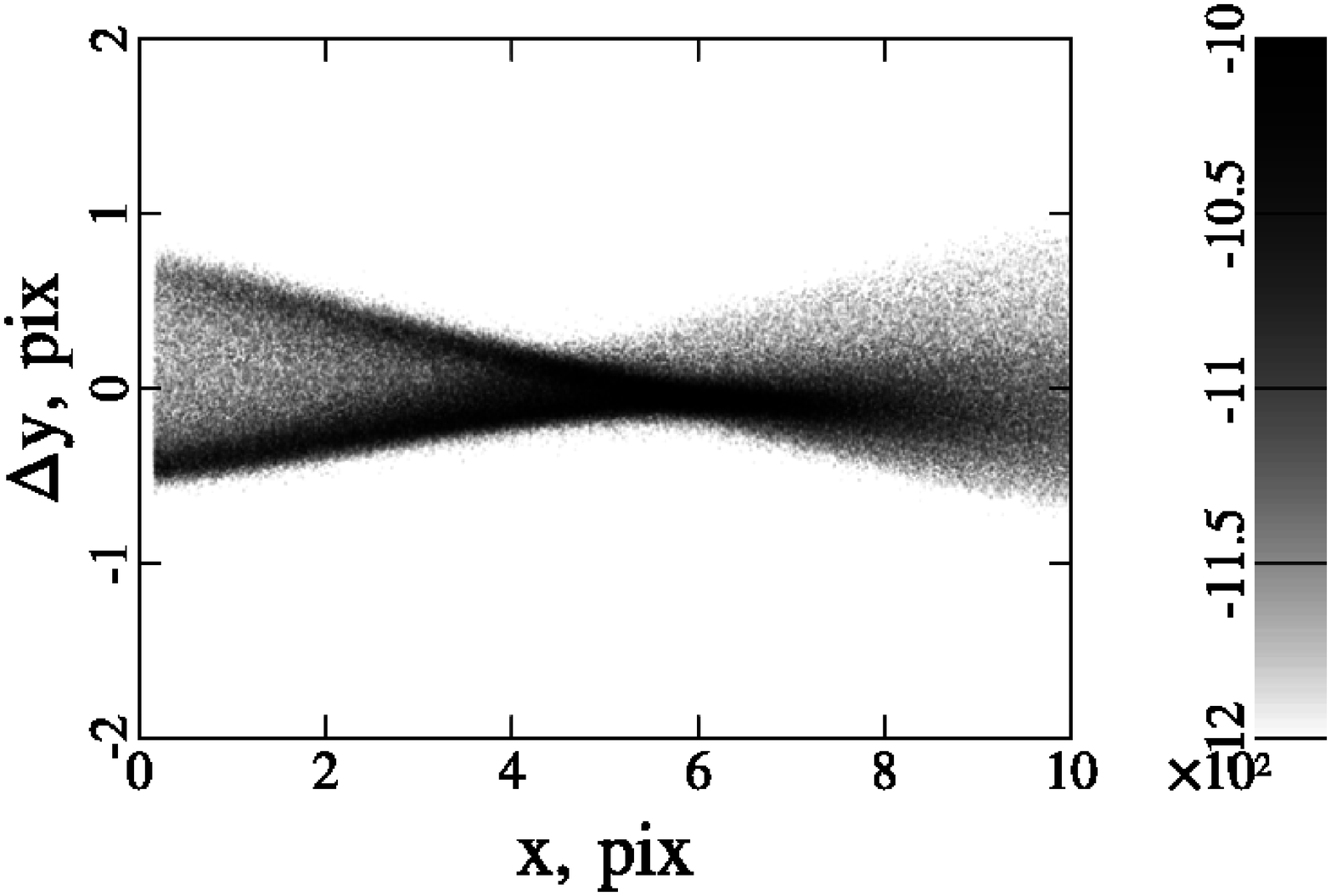} b}
\end{minipage}
\caption{\rm Examples of the dependences of the residual stellar coordinate differences on the magnitude and position on the frame: (a) for the Pulkovo CCD frames (with a scale of 0.950 arcsec/pix) and (b) for the WISE images (in the W1 band, with a scale of 2.758 arcsec/pix).}
\label{fig4}
\end{figure} 

Coma, distortion, or other geometric distortions can be responsible for these differences. They can manifest itself as the dependences of the residuals on coordinates and magnitude. Examples of such dependences are shown in Fig.~\ref{fig4}. Based on the presented graphs, we can state with confidence that there is no magnitude equation for the Pulkovo CCD frames and there is a noticeable and complex distortion for the WISE frames. To calculate the corrections, we divided the working field of the frame into square (or rectangular) cells whose sizes were chosen in such a way that at least 100 differences for each magnitude group (from 10$^{m}$ to 16$^{m}$ with a 1$^{m}$ step) fell into them.

\begin{figure}[h!]
\begin{minipage}[h]{0.5\linewidth}
\center{\includegraphics[width=1\linewidth]{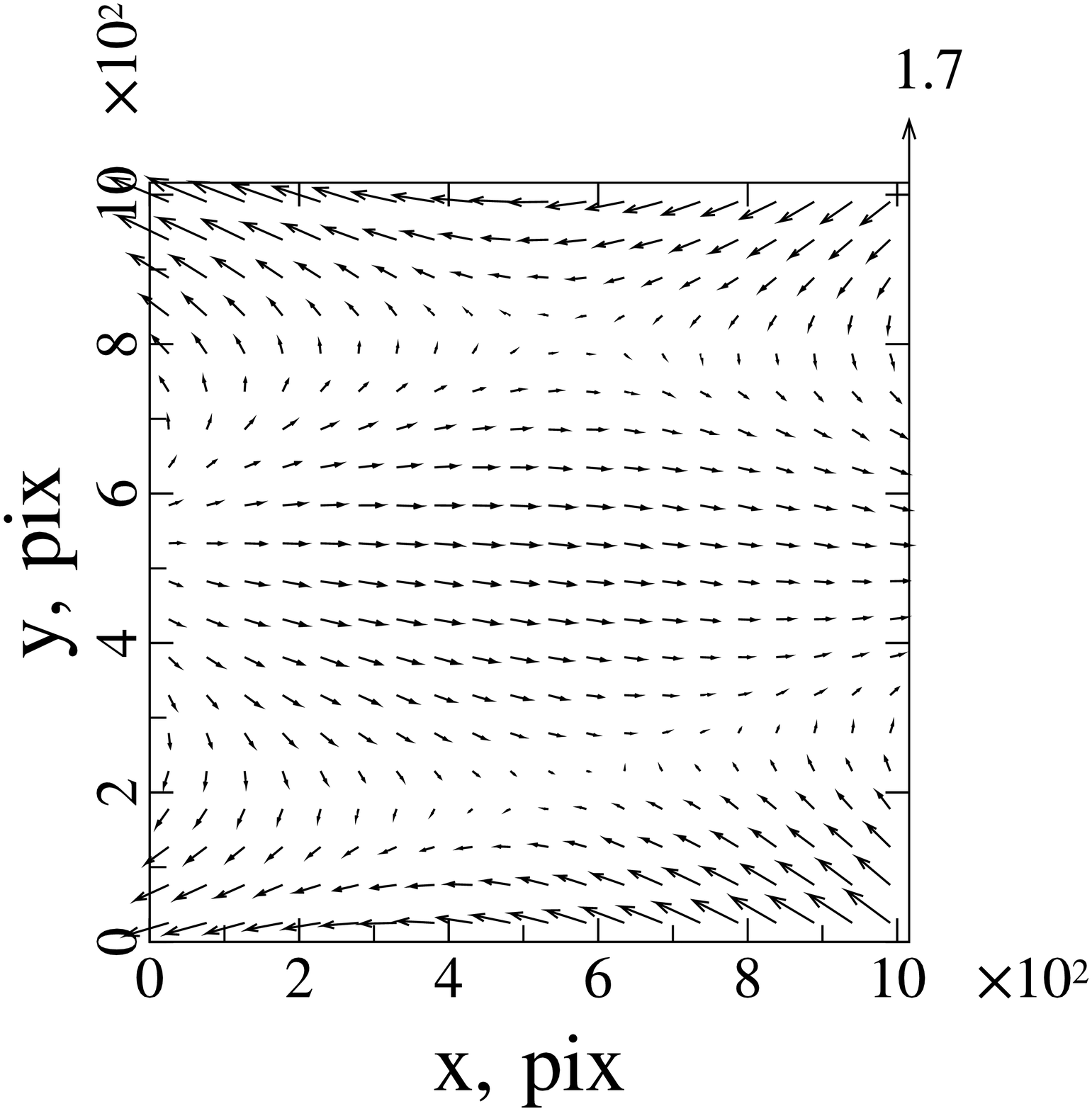} a}
\end{minipage}
\begin{minipage}[h]{0.5\linewidth}
\center{\includegraphics[width=1\linewidth]{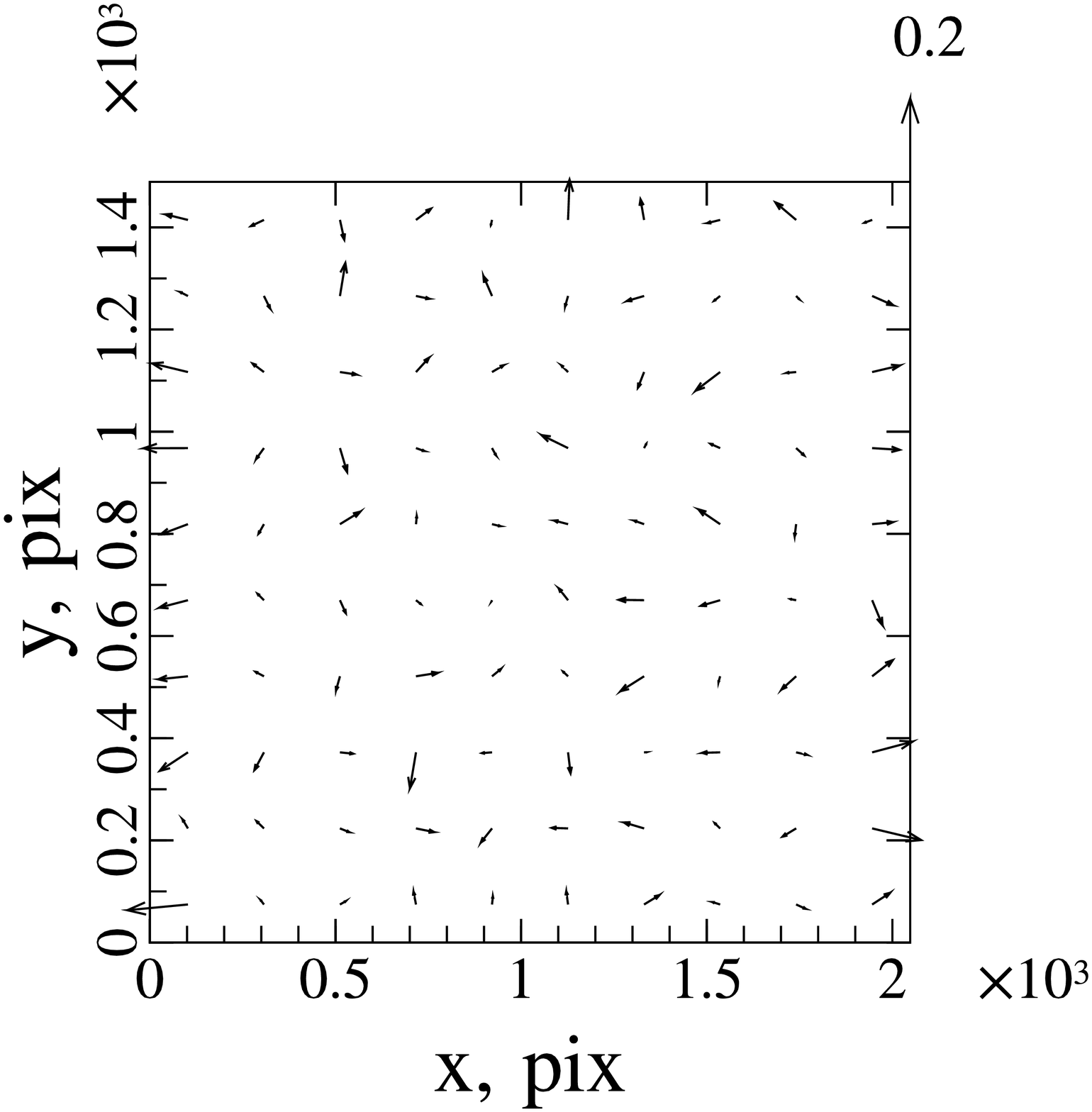} b}
\end{minipage}
\caption{\rm Examples of the vector fields of residual pixel stellar coordinate differences in the range $14^m$ - $16^m$: (a) for all WISE images (in the W1 band, with a scale of 2.758 arcsec/pix) and (b) for all SDSS DR12 images (in the r band, with a scale of 0.396 arcsec/pix). The maximum (in magnitude) vector (its magnitude is given in pix) is shown in the upper right corner.}
\label{fig5}
\end{figure}

As the corrections for the geometric center of a cell, we used the means of all the differences that fell into this cell with the corresponding coordinates of the center and magnitudes. The corrections for any point and the magnitudes were calculated by a bicubic interpolation. Examples of the vector fields of residual differences are shown in Fig.~\ref{fig5}. It can be seen that the influence of a significant distortion of the WISE telescope may well be revealed and properly taken into account. The SDSS frames are free from significant geometric distortions. The systematic errors are tenths or hundredths of a pixel (less than 40–50 mas).

Thus, for all of the surveys used (and for the photometric bands within a survey), we constructed the maps of corrections that allowed us to take into account the most significant systematic errors and to obtain the coordinates of the reference and program stars suitable for deriving the proper motions.

\section*{Revealind $\Delta\mu$ binaries} 
\subsection*{\it Determining the Proper Motions}

For each star, we produced the sets of frames from all possible surveys and found and measured the common stars. It should be noted that quite a few regions are absent in SDSS; in several cases, the image of a program star was measured unsatisfactorily on the WISE frames or the scans of Palomar plates. We chose the stars whose images in the pixel with maximum intensity were characterized by a signal-to-noise ratio $snr>10$ on all frames as the reference ones. Each of these stars has a reliably determined proper motion in the UCAC4 catalog (\citet{Zacharias2013}). We applied the corresponding corrections for the systematic errors and proper motions to the pixel coordinates of the reference stars.

The reference frame was determined in each set. The time at which this frame was taken was closer to the mean epoch of observations calculated from all frames than that for all the remaining ones. For each frame, we calculated the parameters of the linear model of the transition to the reference frame by least squares. As a result, we obtained a set of coordinates in the system of the reference frame for the program star.

In the surveys used, the same region was imaged several times (with different filters). For example, there were at least five frames for the corresponding photometric bands in SDSS. For 2MASS and WISE, the imaging was performed both in different bands and with an overlap (the region of interest to us was present on several frames). As a result, it became possible not only to increase the accuracy of the coordinates by taking their mean but also to estimate the accuracy of their determination. The standard position errors depend on the survey used. They lie within the range 20–40 mas for SDSS and 2MASS and 80–150 mas for the Palomar surveys and WISE; for the Pulkovo series of CCD frames, the internal accuracy of the stellar coordinates on the reference frame is 50–100 mas, depending on the magnitude of a star.

\begin{figure}[h!]
\center{\includegraphics[width=1\linewidth]{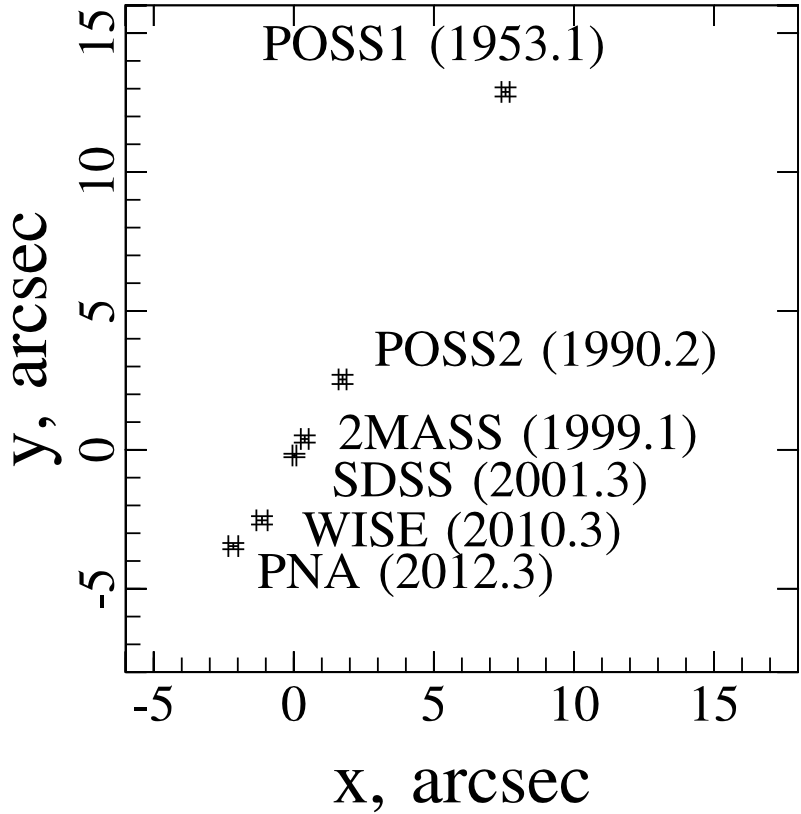}}
\caption{\rm Motion diagram for J0838+4715. The proper motion components are $-158.1\pm2.6$~mas$\cdot$yr$^{-1}$ in right ascension and $-273.1\pm3.1$~mas$\cdot$yr$^{-1}$ in declination. The mean epoch of observations is 1994.3328. $V_{mag} = 15.9$ (PNA stands for the Pulkovo normal astrograph).}
\label{fig6}
\end{figure}

We determined a total of three variants of the proper motion. The first of them is the result of a linear least-squares fitting of the motion over all points. The weights of the points were assigned depending on the standard errors of the stellar coordinates for the corresponding epochs. The first two positions in the set corresponded to the POSS1 (1950s) and POSS2 (mainly 1980s and 1990s) mean epochs. All of the remaining ones correspond to the more recent 2MASS (1998–2002), SDSS (2000–2015), Pulkovo frames (PNA, 2008–2015), and WISE (2010, the cold phase of the mission). The aforesaid is illustrated by the motion diagram for one of the program stars in Fig.~\ref{fig6}. To obtain independent variants, the quasimean proper motion (second variant) was calculated as the POSS2-POSS1 coordinate difference divided by the epoch difference. The last solution was considered as the quasi-instantaneous proper motion (third variant). It was constructed by analogy with the first variant without allowance for the POSS1 and POSS2 data.

\subsection*{\it Calculating the Threshold Value of the Test for Revealing $\Delta\mu$ Binaries}

{\raggedright \citet{Wielen1999} proposed a statistical test that allowed one to decide whether the difference between the quasi-mean ($\mu_{mean}$) and quasi-instantaneous ($\mu_{inst}$) proper motions could be considered statistically significant. In the case of FK5 and Hipparcos data, the proper motions were determined from a large number of observations, and there was no doubt that the distribution of these quantities was Gaussian. The quantity F calculated from the following formula is used to determine the threshold value of the test:\\}
{\centering $F^{2}=(\dfrac{\Delta\mu_{\alpha}cos\delta}{\epsilon_{\mu_{\alpha}}})^{2}+(\dfrac{\Delta\mu_{\delta}}{\epsilon_{\mu_{\delta}}})^{2}$,\\} 
{\raggedright where $\Delta\mu_{\alpha}cos\delta$ and $\Delta\mu_{\delta}$ are the differences of the proper motion components ($\mu_{mean}-\mu_{inst}$), $\epsilon_{\mu_{\alpha}}cos\delta$ and $\epsilon_{\mu_{\delta}}$ are the corresponding mutual standard errors ($\epsilon^{2}_{\mu}=\epsilon^{2}_{\mu_{inst}}+\epsilon^{2}_{\mu_{mean}}$).}

Analysis of the cumulative distribution of this quantity leads to the conclusion that the probability of observing the corresponding proper motion differences for single stars is less than 0.05 for $F > 2.49$. It is such stars that were considered in \citet{Wielen1999} as $\Delta\mu$-binary candidates.

In our case, the number of points from which the inference should be drawn is small (from 4 to 10). Therefore, it was necessary to check whether $F = 2.49$ could be used as the threshold one. For this purpose, we performed numerical simulations of the linear motion of a single star over the celestial sphere with specified position errors spanning a period characteristic for our problem (about 65 years). The points to determine the proper motions were taken at random, given the characteristic implementation times of the surveys used. Next, we performed calculations according to the scheme described in the preceding section, but for a model star.

Different cases were considered: from four (from two Palomar images, Pulkovo frames, and 2MASS frames) to ten points. For each of them, we conducted 100 000 tests. As a result, we constructed the cumulative F distribution and calculated the threshold value for each case.

\begin{figure}[h!]
\begin{minipage}[h]{0.5\linewidth}
\center{\includegraphics[width=1\linewidth]{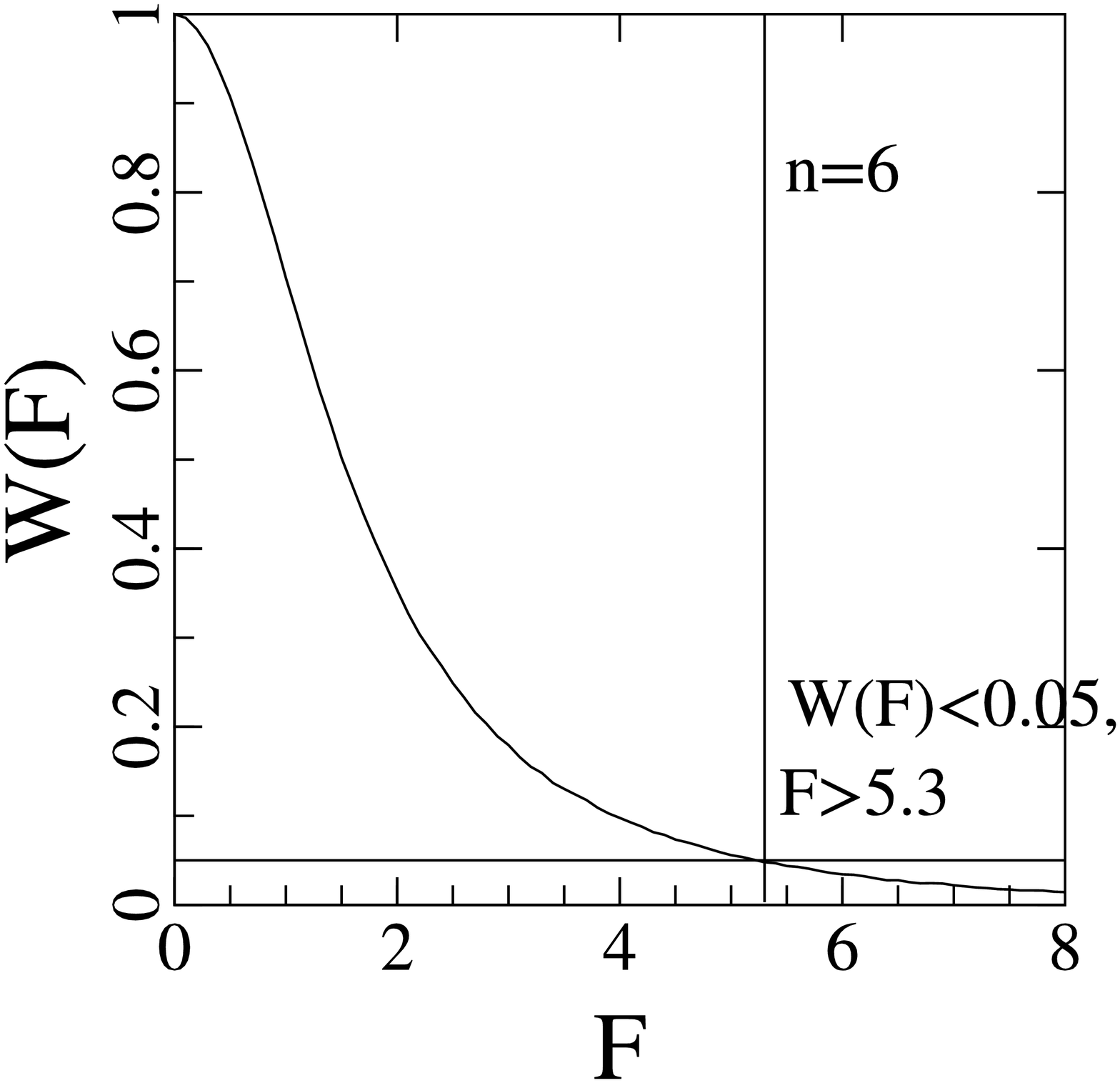} a}
\end{minipage}
\begin{minipage}[h]{0.5\linewidth}
\center{\includegraphics[width=1\linewidth]{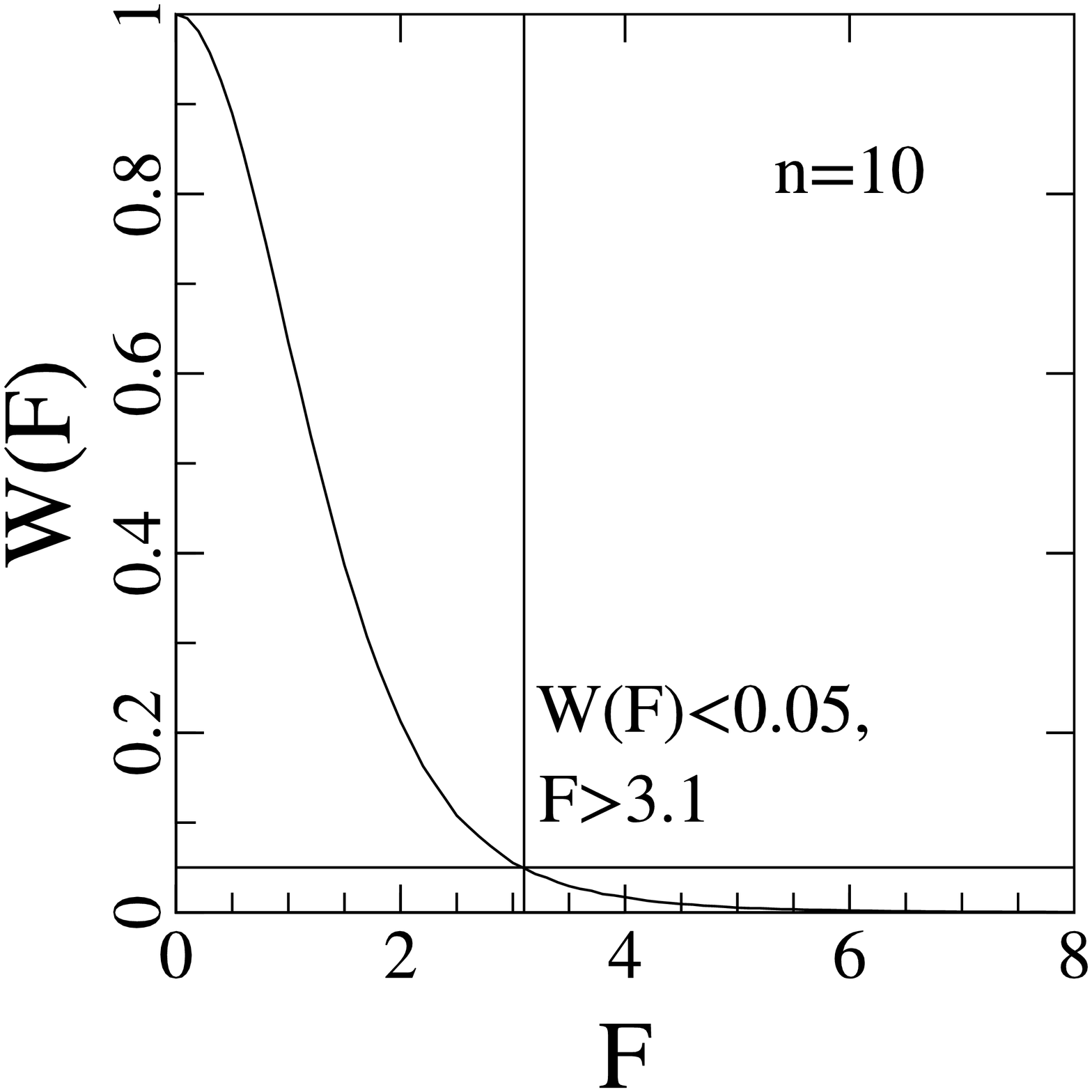} b}
\end{minipage}
\caption{\rm Cumulative F distributions for different numbers of frames used to calculate the proper motions (n = (a) 6 and (b) 10).}
\label{fig7}
\end{figure}

It can be seen from Fig.~\ref{fig7} that $F > 5.3$ for six points, $F > 3.1$ for ten points, and so on. As the number of points increases to 20, the threshold value of this quantity approaches 2.5, which corresponds to the situation considered in \citet{Wielen1999}. The threshold values obtained in this way were used for real stars to reach a conclusion about the pattern of stellar motion over the celestial sphere.

\subsection*{\it General Characteristic of the Derived Proper Motions}

Finally, we processed the frames and plate scans for 1308 stars using the described technique. We failed to reliably measure the positions of 120 stars on the frames from several surveys most often due to the presence of the images of very bright stars nearby. Additional studies are required for these objects.

All our results are summarized in four tables accessible in electronic form~\footnote{http://vizier.u-strasbg.fr/viz-bin/VizieR?-source=J/PAZh/41/896} (as an example, a small part of the main data set is shown in Table~\ref{tab1}). Apart from our calculated proper motions and F-test estimates, additional data (the proper motion differences from our comparison with the LSPM data, the photometric and trigonometric parallaxes, the magnitude estimates for different bands, and membership in the catalogs of binary stars) are given for each star. The service of the Centre de Donnees astronomiques de Strasbourg  \citet{Ochsenbein2000} was used to generate the tables.

The reliability of determining the quasi-instantaneous proper motions (third variant) depends significantly on the epoch difference between the surveys. Therefore, the material was divided into two parts. The first part contains the stars with a corresponding epoch difference of more than 5 years; the second part contains all the remaining ones. There are 944 stars for which no evidence of motion nonlinearity was found and 121 stars manifesting themselves as $\Delta\mu$-binary candidates in the first part. The second group comprises 229 single stars and 14 $\Delta\mu$-binary candidates.

\begin{figure}[h!]
\begin{minipage}[h]{0.5\linewidth}
\center{\includegraphics[width=1\linewidth]{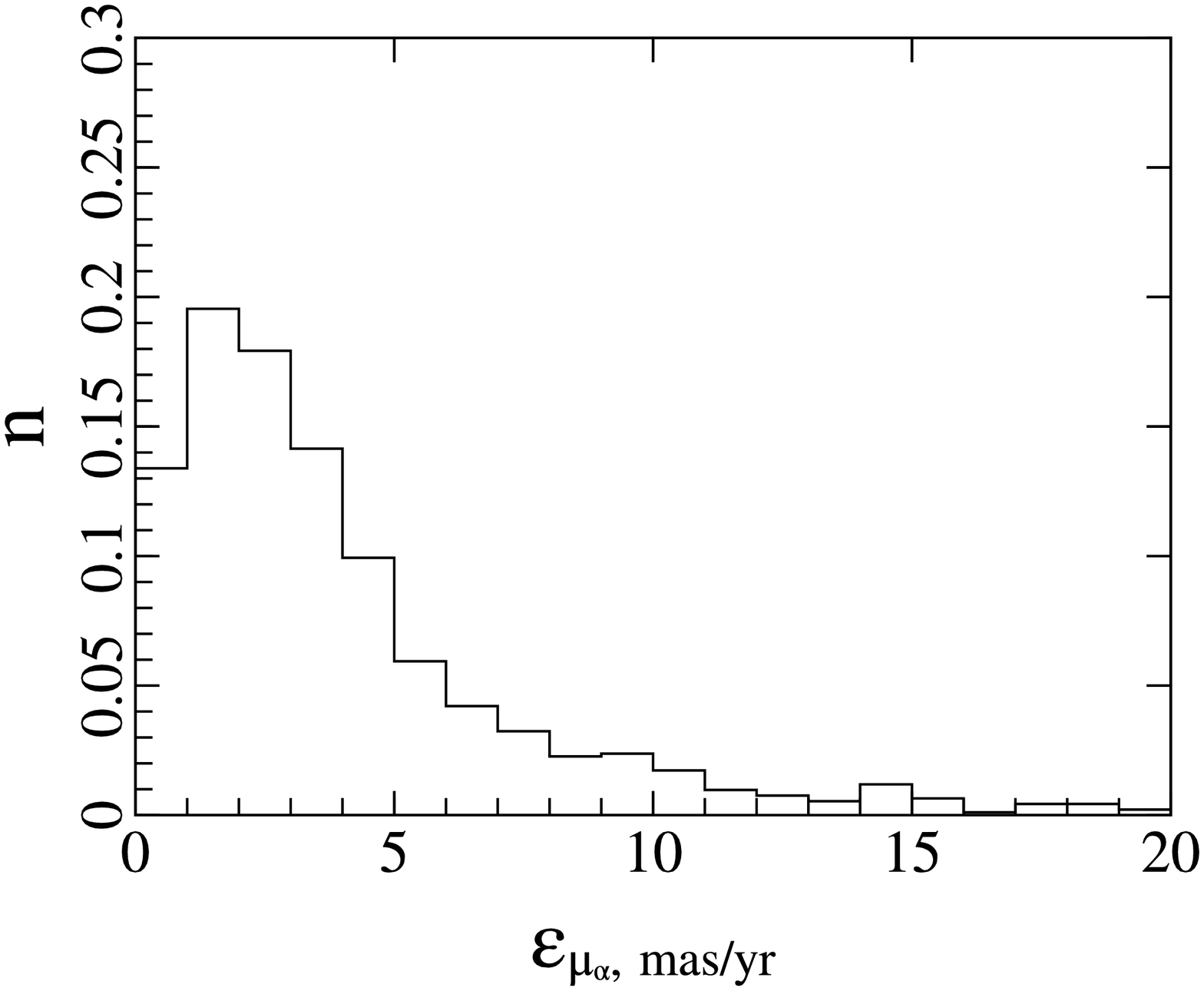} a}
\end{minipage}
\begin{minipage}[h]{0.5\linewidth}
\center{\includegraphics[width=1\linewidth]{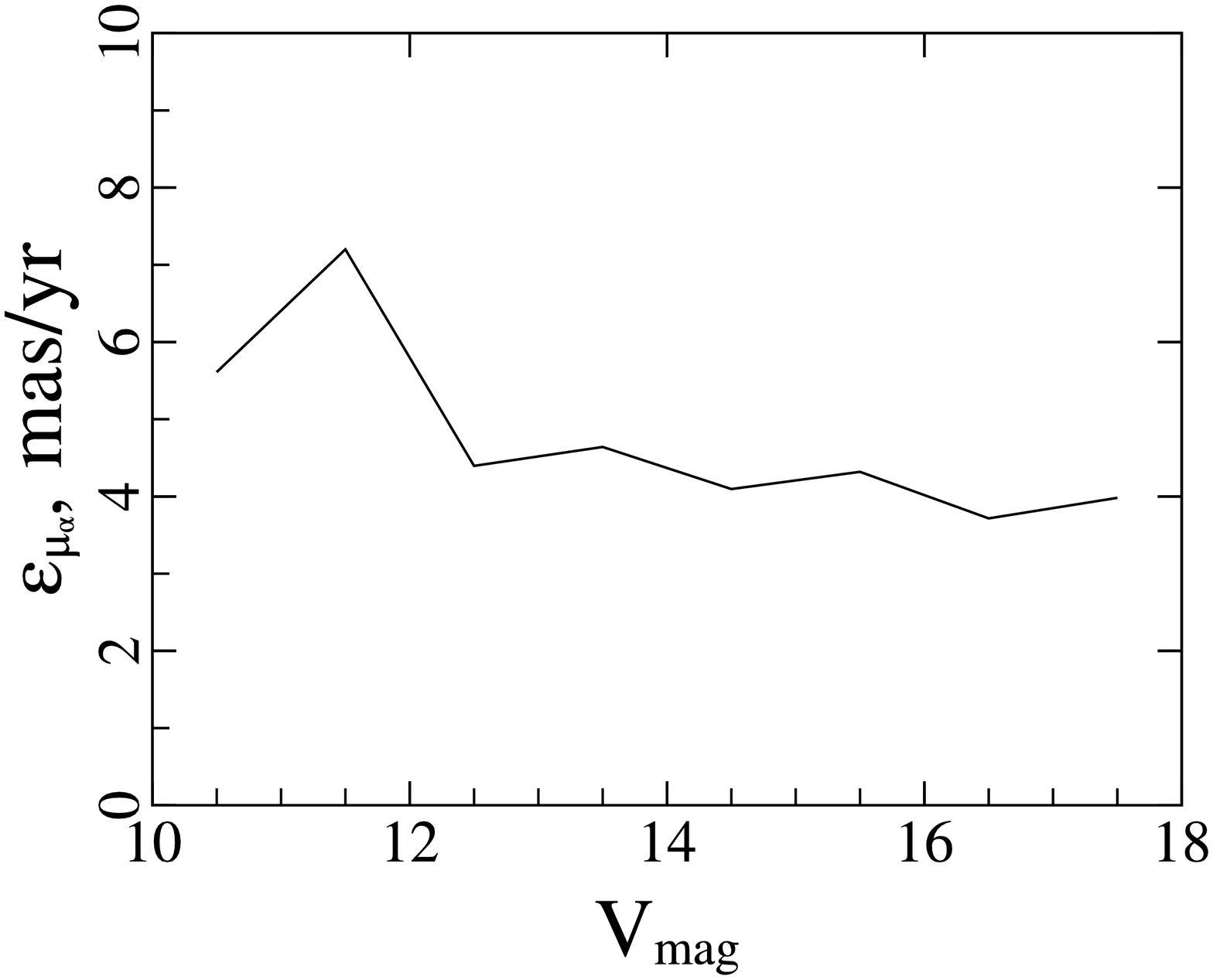} b}
\end{minipage}
\caption{\rm (a) Distribution of proper motion errors and (b) their dependence on magnitude for the first variant of the proper motions, when they were determined from all possible positions from the surveys used.}
\label{fig8}
\end{figure}

For the proper motions constructed from all surveys (first variant), the mean accuracy is slightly lower: 4 mas$\cdot$yr$^{-1}$ in both coordinates. To get a more comprehensive idea of the quality of the proper motions, Fig.~\ref{fig8} presents the corresponding histogram and the magnitude dependence of the proper motion errors (for the first variant of the proper motions). For the overwhelming majority of stars, the proper motion errors do not exceed 10 mas$\cdot$yr$^{-1}$.

The quasi-mean proper motions are characterized by a mean accuracy of about 5–10 mas$\cdot$yr$^{-1}$. For the quasi-instantaneous proper motions, it is 10–20 mas$\cdot$yr$^{-1}$ (it depends significantly on the availability of SDSS frames and the magnitude of a star).

\begin{figure}[h!]
\begin{minipage}[h]{0.5\linewidth}
\center{\includegraphics[width=1\linewidth]{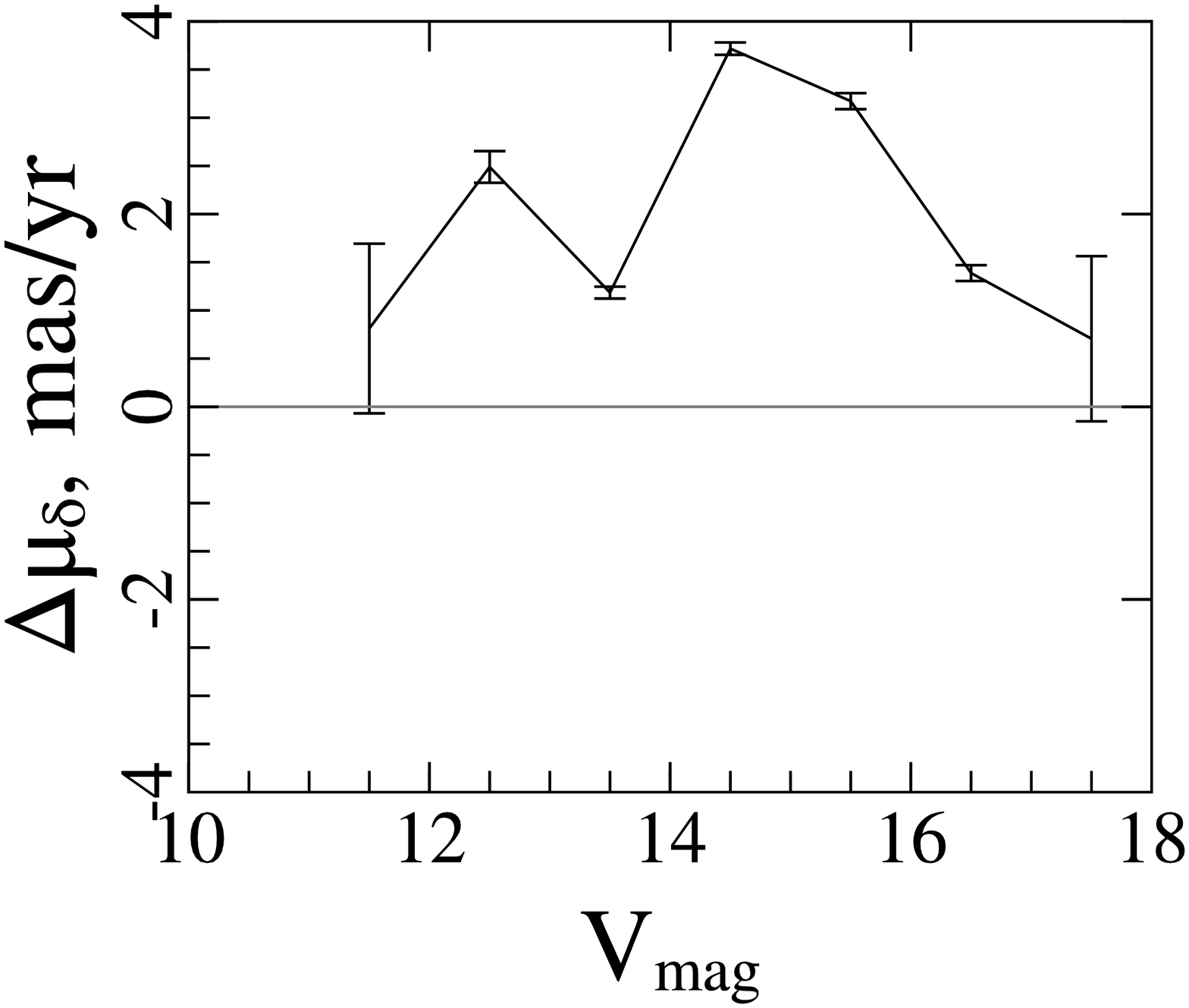} a}
\end{minipage}
\begin{minipage}[h]{0.5\linewidth}
\center{\includegraphics[width=1\linewidth]{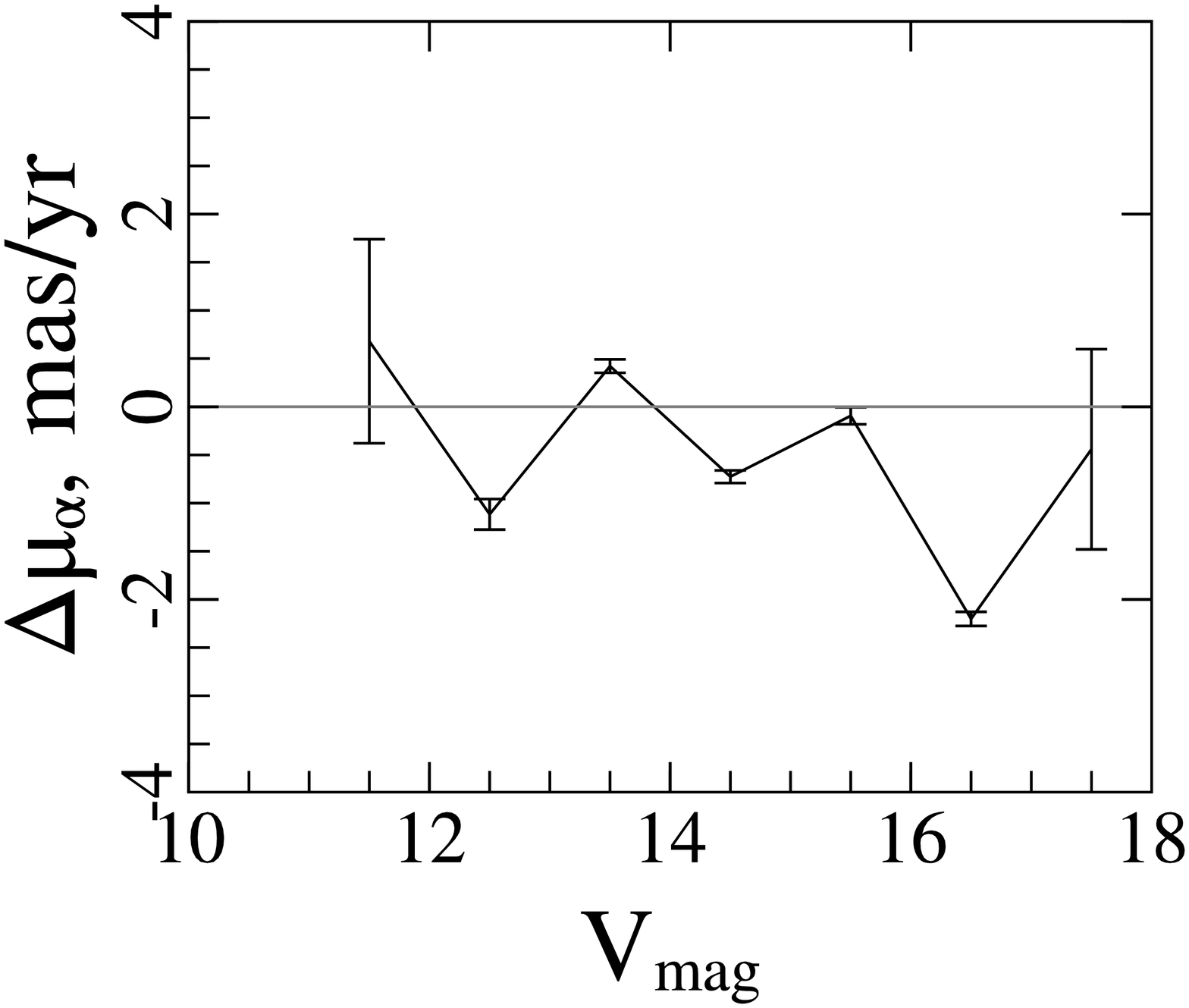} b}
\end{minipage}
\caption{\rm Magnitude dependence of the proper motion differences $\mu - \mu_{LSPM}$ ($\mu$ is the proper motion calculated from all frames).}
\label{fig9}
\end{figure}

The existence of systematic differences between our set of proper motions and the stellar proper motions in the LSPM catalog is demonstrated by Fig.~\ref{fig9}. These differences are particularly prominent for the proper motions in declination (the differences can reach 2–4 mas$\cdot$yr$^{-1}$).

A peculiarity of the LSPM catalog is a nonstandard approach to absolutizing the proper motions. In the period of its production, the faint reference stars had not yet been provided with reliable proper motions. Therefore, this procedure was performed using bright stars from the Tycho-2 catalog whose images on the Palomar survey plates were either overexposed or close to this. The extrapolation of the absolutizing corrections to the region of fainter (by 3$^{m}$ - 5$^{m}$) stars was not checked in any way, which was probably responsible for the observed systematic differences.

\subsection*{\it Verification of the Revealed $\Delta\mu$-Binary Candidates}

Only 121 objects show evidence for nonlinearity of their motion over the celestial sphere, classifying such stars as $\Delta\mu$ binaries. In our view, these objects are of interest for further studies to confirm their binarity and to determine their orbits and masses, for example, through speckle interferometry with large telescopes. For 14 stars from the second group for which the quasi-instantaneous proper motion differs noticeably from the quasi-mean one, the conclusion about their n belonging to $\Delta\mu$ binaries seems very unreliable due to a small epoch difference in present-day sky surveys. Therefore, our subsequent analysis concerns only the
stars from the first group.

Ten stars from our list of $\Delta\mu$ binaries enter into the WDS catalog (\citet{Mason2001}), i.e., they are visual double stars (these account for $9\%$ of the total number of $\Delta\mu$ binaries). Most of them are wide pairs with a common proper motion (the angular separation between the components is tens of arseconds and, in several cases, arcminutes). This result is a weak confirmation of the efficiency of the technique for revealing $\Delta\mu$ binaries. In the list of 944 single stars, 58 enter into WDS ($6\%$); therefore, the fraction of known visual double stars in the list of $\Delta\mu$ binaries cannot be said to be significantly larger.

Some of the Pulkovo program stars have the trigonometric parallaxes derived from ground-based observations. The proper motions, along with the parallaxes, are known to be also determined with a high accuracy during such observations over a short interval (3–5 years). Therefore, they may be considered as the quasi-instantaneous proper motions, and $F_{\pi}$ can be estimated by comparing them with the proper motions determined from our results.

\begin{figure}[h!]
\center{\includegraphics[width=1\linewidth]{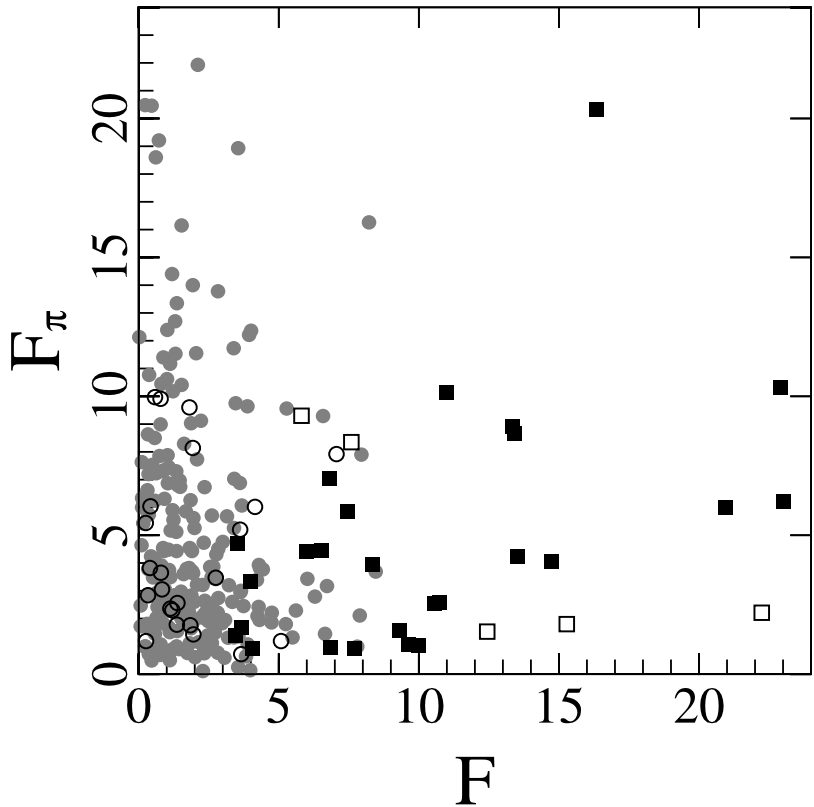}}
\caption{\rm Distribution of stars in the $F-F_{\pi}$ plane. $F$ are the values of the test obtained by analyzing the CCD frames and scans from the surveys. We determined $F_{\pi}$ based on the proper motions from this paper and those obtained when various parallactic programs were implemented (the circles and squares denote the single stars and $\Delta\mu$ binaries, respectively; the open symbols correspond to the stars from the WDS catalog).}
\label{fig10}
\end{figure}

Our list of $\Delta\mu$ binaries contains 33 stars that enter into the parallactic programs. Eighteen of them have $F_{\pi} > 3$ (and five stars enter into WDS). The correlation between the independent sets of $F$ and $F_{\pi}$ can be judged in more detail from Fig. ~\ref{fig10}. With some reservation, we can talk about the existence of a correlation between $F$ and $F_{\pi}$ for $\Delta\mu$ binaries. In the case of stars considered as single ones, the evidence of a correlation is not so clear. The positions of visual double stars from the WDS catalog are marked in Fig. 10 by open symbols. They are present among both $\Delta\mu$ binaries and single stars. And this is quite natural. For wide pairs with long revolution periods (hundreds or thousands of years), which are quite a few in our material, it is fairly difficult to detect the orbital motion based on the series of observations used.

\begin{figure}[h!]
\center{\includegraphics[width=1\linewidth]{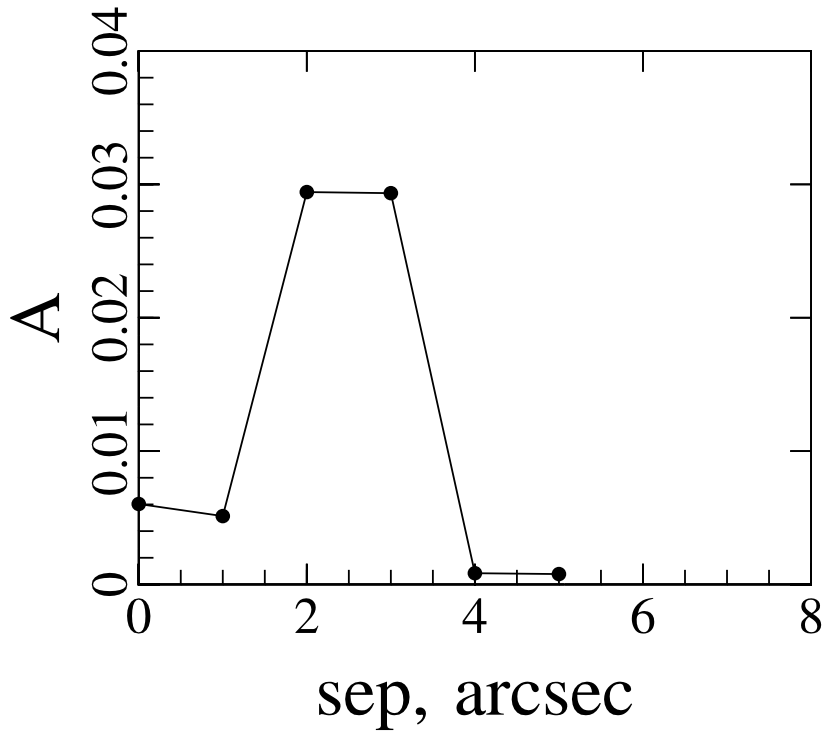}}
\caption{\rm Simulation of the dependence of the asymmetry of interacting stellar images on the angular separation between the stars (the primary component is $14.1^m$; the secondary star is $15.3^m$). The results of the shapelet decompositions of stars on the SDSS frames were taken as the basis.}
\label{fig11}
\end{figure}

For binary systems of dwarf stars close to the Sun, the angular separations between their components can be one arcsecond or more. Therefore, binarity can be established simply by analyzing the images of stars on the CCD frames. The shapelet decomposition used to determine the pixel coordinates of stars also allows the image asymmetry to be estimated. To understand how reliable this approach is, we performed numerical simulations of the images of such pairs. For SDSS, which has the best resolution among all of the surveys used, we constructed the model interacting images for binary systems with typical magnitudes of their components at various angular separations between them based on the results of fitting the images of single stars. These interacting images were then again subjected to the shapelet decomposition. An example of the dependence of the asymmetry on angular separation for a pair with component magnitudes of 14.1$^{m}$ and 15.3$^{m}$ is shown in Fig.~\ref{fig11}. Analysis of this graph suggests that the image asymmetry estimates are most effective for angular separations of 2–4 arcsec. The asymmetry is also noticeable for smaller separations, but additional studies are required to estimate the reliability of the approach under consideration for such close pairs. Given the image quality and the scales of the CCD frames and scans from the remaining surveys used, it can be concluded that the image asymmetry estimates are not enough to establish binarity. Nevertheless, the asymmetry is highly informative for analysis, along with the F-test estimates. Therefore, the final table with the results includes the mean values of the asymmetry based on all surveys; the maximum values of the asymmetry based on the entire sample for a given star are provided with an indication of the corresponding survey (Table~\ref{tab1}). Curiously, the mean stellar image asymmetry is 0.039 for the sample of $\Delta\mu$ binaries and 0.025 for single stars.

The parameters of binary systems from the WDS catalog (component separation, position angles, component magnitudes) were included in the electronic versions of the tables with the results. For the $\Delta\mu$ binaries J0027+5330N, J0514+4431N, J0517+4550, J1614+6038, and J1938+3512E, the angular separations between their components lie within the range from fractions of an arcsecond to 4 . The star J1938+3512E enters into SDSS and has an asymmetry of 0.086, which is appreciably higher than its mean values for single stars of the same magnitude in this survey (usually from 0 to 0.04). These stars can be singled out as examples of the successful detection of nonlinear motion over the celestial sphere based on comparison of the quasi-mean and quasi-instantaneous proper motions.

\begin{table}
\fontsize{8}{8}\selectfont
\vspace{6mm}
\raggedright
\caption{\rm Fragments of the table for $\Delta\mu$-binary candidates}
\label{tab1}
\vspace{5mm}

\begin{tabular}{c|c|c|c|c|c|c|c|c|c} \hline 
	& \multicolumn{4}{|c|}{Quasi-mean $\mu$ }&\multicolumn{4}{|c|}{Quasi-instantaneous $\mu$}&\\
\cline{2-9}
LSPM&$\mu_\alpha$&$\mu_\delta$&$\varepsilon_{\mu_\alpha}$&$\varepsilon_{\mu_\delta}$
&$\mu_\alpha$&$\mu_\delta$&$\varepsilon_{\mu_\alpha}$&$\varepsilon_{\mu_\delta}$&T\\
\cline{2-9}
	   &\multicolumn{8}{|c|}{mas$\cdot$yr$^{-1}$}&\\
\hline
J1756+5132 &   -526.8&    205.3&   5.3&   5.2&   -457.7&    -19.4&  17.1&  18.2&   2007.0035\\
J1844+6511 &    -16.5&    356.4&   7.1&   5.4&     30.0&    289.0&   0.3&   3.7&   2005.4087\\
J1908+3216 &   -238.5&   -227.0&   6.8&   6.3&   -200.9&   -215.4&   0.5&   9.5&   2006.9105\\
J1931+4115 &   -249.7&   -119.4&  15.6&  14.1&   -329.5&     -1.4&   7.9&   4.4&   2007.7614\\
J1931+6843 &    256.2&    461.7&   6.3&   4.9&    228.9&    349.9&   5.8&   0.4&   2006.9060\\
J1938+3512E&     -2.1&    814.7&  11.4&  12.2&   -123.3&    714.9&   6.1&   0.3&   2004.6930\\
\hline\hline
  LSPM     &   F    & $n$ &  $V_{mag}$&{\small$V-J$}&$\rho$,mas& $\rho/\varepsilon$&$\left\langle A \right\rangle$&$A_{max}$&Survey\\ 
\hline
J1756+5132 &   12.50&   4 &  15.96&   3.18 &     &     & 0.040&  0.080&poss1\\
J1844+6511 &   12.23&   5 &  16.46&   4.08 &     &     & 0.026&  0.060&poss2\\
J1908+3216 &    5.57&   6 &  11.83&   3.92 &     &     & 0.019&  0.045&~wise\\
J1931+4115 &    9.18&   7 &  12.20&   2.12 & 1241&  5.7& 0.032&  0.080&poss2\\
J1931+6843 &   22.90&   5 &  14.67&   4.65 &     &     & 0.027&  0.050&poss1\\
J1938+3512E&   12.44&   5 &  14.75&   2.98 &     &     & 0.038&  0.086&~sdss\\
\hline
\end{tabular}

\raggedright
\vspace{5mm}
{\footnotesize
$\mu_\alpha$,$\mu_\delta$,$\varepsilon_{\mu_\alpha}$,$\varepsilon_{\mu_\delta}$
 are the proper motion components and their standard errors.
T is the mean epoch for the quasi-instantaneous proper motion.
F is the value of the test for “nonlinearity” of the stellar motion.
$n$ is the number of surveys used.
$V_{mag}$ and {\small$V-J$} are the magnitude and color index.
$\rho$ is the angular separation between the optical and infrared positions of the photocenter.
$\rho/\varepsilon$ is the ratio of $\rho$ to the standart error ($\varepsilon$) of the stellar coordinates.
$\rho$ and $\rho/\varepsilon$ are given only if $\rho/\varepsilon>5$.
$\left\langle A \right\rangle$ is the mean stellar image asymmetry for all surveys.
$A_{max}$ is the maximum stellar image asymmetry and the name of the corresponding survey.
The star J1938+3512E enters into the WDS catalog (19389+3512, angular separation $\rho=4''$, position angle $PA=77^{\circ}$, primary magnitude - $mag_p = 15.3$, secondary magnitude $mag_s = 16.00$).}\\

\end{table}

The positions of the image photocenters for the binary systems under consideration can differ, depending on the effective wavelength in the case of a noticeable difference between the effective temperatures of the components (for example, for M dwarf + brown dwarf binary systems). If the surveys were performed simultaneously, then this fact could be easily detected by comparing the stellar coordinates and identifying noticeable shifts of the image centers from the optical surveys to the infrared ones. The WISE space survey is more recent and deeper than 2MASS. Astrometrically, however, 2MASS has a considerably higher quality (the astrometric reduction errors can reach 100–200 mas for WISE and, as a rule, less than 50 mas for 2MASS). Therefore, if the number of stellar positions in the optical surveys (POSS1, POSS2, SDSS, PNA) was at least four, then the proper motion was calculated only from the optical survey data. Using this proper motion, we calculated the optical stellar position for the 2MASS epoch. As a result, the shift ($\rho$) between the optical and 2MASS positions could be estimated. If $\rho > 5\sigma$, then the shift and its ratio to the mutual position error are given in the final table. Consider, for example, J1931+4115 (Table~\ref{tab1}). The shift for this star exceeds $1''$. The image asymmetry for this star is close to the mean for the sample of $\Delta\mu$ binaries and larger than the mean for single stars. Such a joint analysis of the set of tests leads us to conclude that the proposed technique of searching for $\Delta\mu$ binaries is efficient for a certain class of binary systems among the low-luminosity stars.

\begin{figure}[h!]
\begin{minipage}[h]{0.5\linewidth}
\center{\includegraphics[width=1\linewidth]{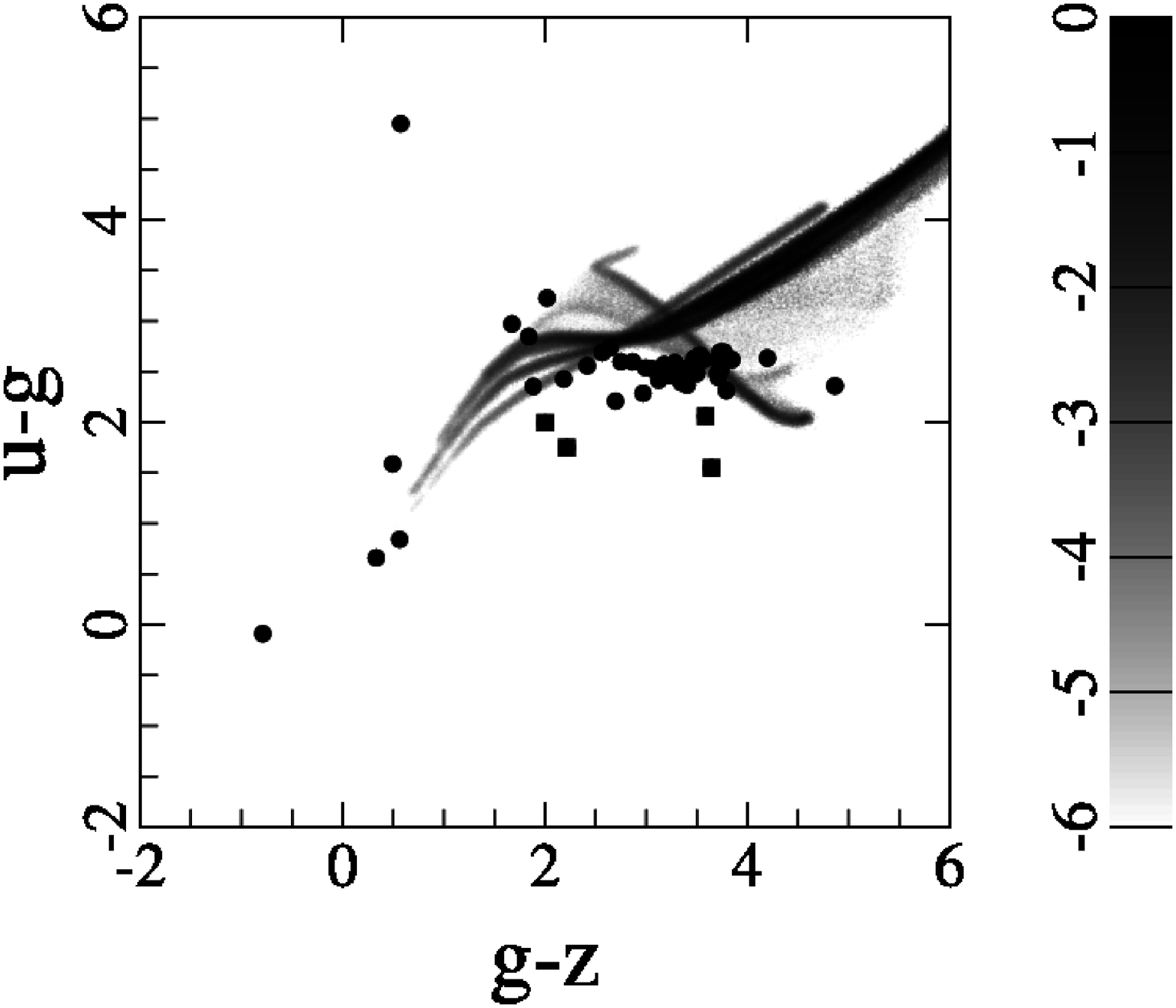} a}
\end{minipage}
\begin{minipage}[h]{0.5\linewidth}
\center{\includegraphics[width=1\linewidth]{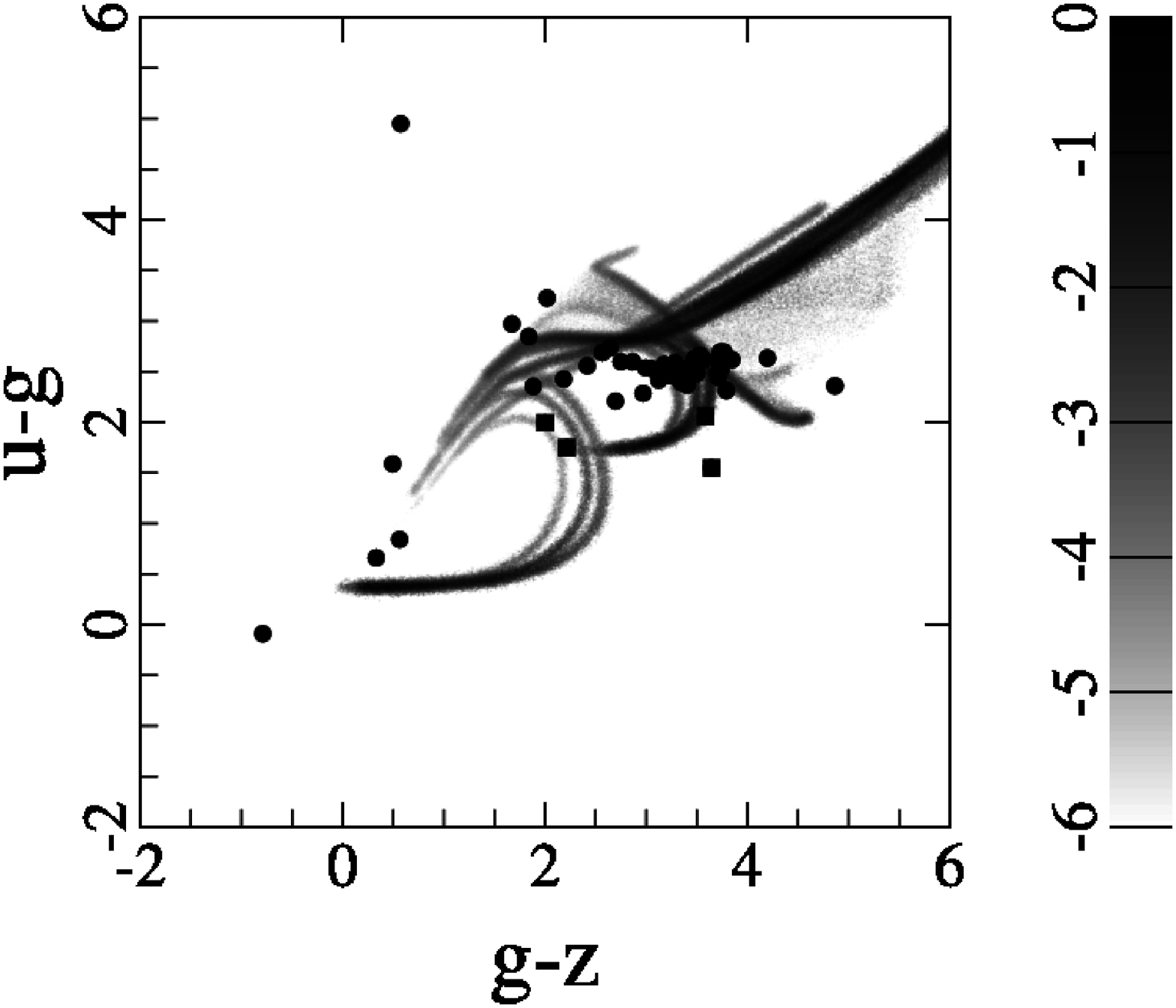} b}
\end{minipage}
\caption{\rm Positions of the $\Delta\mu$-binary candidates on the color–color $(g-z)~-~(u-g)$ diagram. The theoretical sequences for stars in the range from 0.1 to 0.7 $M_\odot$ constructed from the Padova isochrones and the Besan{\c c}on model of the Galaxy for single stars (a) and when adding M dwarf + white dwarf binary systems (b) are shown. The white dwarf masses are 0.3 and 0.6 $M_\odot$ (the ages are 0.6 and 1 Gyr, respectively). The black squares correspond to the stars that may turn out to be M dwarf + white dwarf binary systems. The scale on the right shows the natural logarithm of the relative density of model stars on the diagram.}
\label{fig12}
\end{figure}

Another possibility of verifying the detection of $\Delta\mu$ binaries is to analyze the photometric data. The most accurate available magnitude estimates for the revealed stars are present in SDSS (these are the $u, g, r, i$, and $z$ magnitudes). Using the Besan{\c c}on model of the Galaxy (\citet{Robin2003}) and the Padova isochrones~\footnote{http://stev.oapd.inaf.it/cgi-bin/cmd} (\citet{Bressan2012}; \citet{Chen2014}; \citet{Tang2015}), we constructed the color–color $(g-z)~-~(u-g)$ diagrams for stars with masses from 0.1 to 0.7 $M_\odot$. The technique for constructing the diagrams is identical to the technique used previously in implementing the Pulkovo program of research on stars with large proper motions (\citet{Khovritchev2013}). To understand how the mixing of the colors of unresolved stellar pairs affects the distribution of the density with which the stars fill the $(g-z)~-~(u-g)$ space, we constructed the diagrams both for single stars and by taking into account various combinations of binary components depending on the masses and ages (the fraction of binary stars was assumed to be 0.5). The positions of the revealed $\Delta\mu$-binary candidates on the diagrams for single stars (Fig.~\ref{fig12}a) and with allowance made for the fraction of binary stars (Fig.~\ref{fig12}b).

The method of searching for $\Delta\mu$ binaries is maximally efficient at a large difference between the positions of the center of mass and the photocenter of the binary system. Therefore, searching for the M dwarf + white dwarf pairs for which the component luminosity ratio is noticeably larger than their mass ratio arouses great interest. Simulations show that for most combinations of M dwarfs with different masses, the deviations from the main set of points on the $(g-z)~-~(u-g)$ diagram are imperceptible, while the M dwarf + white dwarf pairs can give a noticeable effect. To estimate it clearly, we took the models of white dwarfs with masses of 0.3 and 0.6 $M_\odot$ and ages of 0.6 and 1 Gyr (\citet{Holberg2006}; \citet{Bergeron2011}; \citet{Kowalski2006}; \citet{Tremblay2011})~\footnote{http://www.astro.umontreal.ca/$\sim$bergeron/CoolingModels} as an example. The mixing of the colors of these white dwarfs with the colors of M dwarfs with different masses and corresponding ages (given that the white dwarfs have passed the main-sequence phase of evolution) is responsible for the difference in the form of the diagrams on the left and right panels of Fig.~\ref{fig12}. Analysis of the deviations of $\Delta\mu$ binaries from the main set of points in this figure gives ground to conclude that four stars (J0656+3827, J0838+3940, J1229+5332, and J2330+4639 denoted by the squares) may be considered as possible pairs of this type.

\begin{table}[h!]
\fontsize{10}{12}\selectfont
\raggedright
\caption{\rm Probable M dwarf + white dwarf pairs}
\label{tab2}
\vspace{5mm}
\begin{tabular}{c|c|c|c|c|c} \hline 	
LSPM
       & $\alpha_{J2000}$& $\delta_{J2000}$&$\mu_\alpha$&$\mu_\delta$&T\\
\cline{2-5}	   
	   & h~m~s         & $\deg~ '~'' $  & \multicolumn{2}{|c|}{mas$\cdot$yr$^{-1}$} \\	   
\hline	   
   
J0656+3827 &06:56:15.9901&+38:27:45.954&    320.5&   -105.9& 1995.8213\\
           &           61&           68&      4.5&      5.4&          \\   
J0838+3940 &08:38:16.7026&+39:40:49.250&    157.1&    107.6& 1994.1300\\
           &          102&          157&      5.5&      8.9&          \\
J1229+5332 &12:29:15.5864&+53:32:43.762&  -1231.7&    139.8& 1991.6092\\
           &           38&            7&      2.8&      0.9&          \\
J2330+4639 &23:30:41.5392&+46:39:56.627&    483.7&    -90.7& 1995.3407\\
           &          124&          198&      6.6&     14.8&          \\
\hline
\end{tabular}
\vspace{5mm}
\\
{\footnotesize        
Here, $\mu_\alpha$ and $\mu_\delta$ are the proper motion components calculated from all surveys.\\
The second row contains the standard position (mas)and proper motion (mas$\cdot$yr$^{-1}$) errors for each star.}\\

\vspace{5mm}\begin{tabular}{c|c|c|c|c|c|c|c|c|c} \hline
      & \multicolumn{4}{|c|}{Quasi-mean $\mu$ }&\multicolumn{4}{|c|}{Quasi-instantaneous $\mu$}&  \\
\cline{2-9}
LSPM 	   &   $\mu_\alpha$&$\mu_\delta$&$\varepsilon_{\mu_\alpha}$&$\varepsilon_{\mu_\delta}$
	   &   $\mu_\alpha$&$\mu_\delta$&$\varepsilon_{\mu_\alpha}$&$\varepsilon_{\mu_\delta}$
	   &   T\\
\cline{2-9}	   
	   &\multicolumn{8}{|c|}{mas$\cdot$yr$^{-1}$}&\\
\hline

J0656+3827 &    313.9&   -115.6&   6.4&   6.6&    353.9&    -80.9&   3.6&   2.1&   2006.7607\\
J0838+3940 &    150.7&    117.7&   5.0&   3.7&    159.7&     63.3&   3.7&   4.0&   2007.8060\\
J1229+5332 &  -1227.4&     25.3&   5.1&   3.9&  -1209.1&    137.4&   2.2&   3.3&   2003.8453\\
J2330+4639 &   1317.4&    836.3&   4.6&   5.6&    512.5&      8.4&  71.0&  99.9&   2005.9775\\
\hline
\end{tabular}

\vspace{5mm}\begin{tabular}{c|c|c|c|c|c|c|c|c} \hline

LSPM &   F    &  n  &  $V_{mag}$&{\small$V-J$}&$\pi_{ph}$,mas&$<A>$& $A_{max}$&Survey\\
\hline

J0656+3827&    7.38& 6 &  14.43&   4.08&  29&0.039 &0.193& sdss\\
J0838+3940&   10.16& 5 &  14.58&   4.08&  27&0.019 &0.030&poss2\\
J1229+5332&   22.24& 5 &  14.21&   4.23&  38&0.041 &0.140& sdss\\
J2330+4639&   14.02& 6 &  13.78&   3.81&  29&0.066 &0.244& sdss\\
\end{tabular}         	   
\\
\vspace{5mm}
{\footnotesize The trigonometric parallaxes $\pi_{yale}=39.9\pm1.0$~mas and $\pi_{MEarth}=44.8\pm3.6$~mas were determined for J1229+5332. This star is presented in WDS (12294+5333, angular separation $\rho=21.4''$, position angle $PA=354^{\circ}$, primary magnitude $mag_p = 13.74$, secondary magnitude $mag_s = 18.20$).}\\ 
\end{table}
\clearpage
Detailed information about these four stars is given in Table~\ref{tab2}. Three of the four stars are characterized by large ($>0.1$) stellar image asymmetries obtained by analyzing the SDSS CCD frames. The star J2330+4639, which enters into WDS as a comparatively wide pair (the component separation is 21.4 arcsec), attracts our attention. For this star, the quasi-instantaneous and quasi-mean proper motions differ greatly at a very large asymmetry (0.244). This can be explained as a consequence of the interaction of the images for two closely spaced stars, which could be responsible for the systematic error of the stellar coordinates in one of the present-day surveys. When analyzing this information, it should be kept in mind that in crowded stellar fields (when tens of stars are observed on a small (several arcmin) area of the frame), the rapidly flying stars close to the Sun can be very close to the background stars on the frames of individual surveys. Therefore, the formation of interacting images is possible in a situation where a binary star is out of the question. However, this is rather difficult to check so far, because the surveys used have different limiting magnitudes.

\section*{Conclusions}

From 2008 to 2015, we carried out astrometric observations of 1428 stars from the list of 1972 objects within the framework of the Pulkovo program of research on stars with large proper motions. For 1308 of them, we downloaded the scans of Palomar plates and the CCD frames from the 2MASS, SDSS, and WISE digital sky surveys.

As a result of our measurements of the common reference stars and once the coordinates have been corrected for the systematic errors, we obtained the proper motions of the program stars in the HCRF/UCAC4 system at an accuracy level of 4 mas$\cdot$yr$^{-1}$. This material can be in demand after the appearance of the first GAIA release to compare the proper motions and to quickly reveal binary stars.

Based on the analyzed material, we constructed the quasi-instantaneous and quasi-mean proper motions whose comparison revealed 121 stars with evidence for nonlinearity of their motion over the celestial sphere ($\Delta\mu$ binaries). The consistency of the applied technique is partly confirmed by comparing the derived proper motions with the data of parallactic programs (Yale, MEarth, and Pulkovo) and analyzing the asymmetry of stellar images on the CCD frames from the surveys used. Differences between the optical and infrared positions of the photocenter calculated for the 2MASS epoch are noticeable for a number of stars. Our study of the positions of the revealed $\Delta\mu$-binary candidates on a color–color diagram suggests that the stars J0656+3827, J0838+3940, J1229+5332, and J2330+4639 may be M dwarf + white dwarf pairs.

\section*{Acknowledgments}

We wish to thank all the observers of the Pulkovo normal astrograph who participated in implementing the observational program. Our work was supported in part by Program P-41 of the Presidium of the Russian Academy of Sciences (“Transitional and Explosive Processes in Astrophysics”). We are grateful to I.V. Samsonov and the “Intellect” Center~\footnote{http://intellect.lokos.net/} for the access to the computer server that allowed us to download the data and to organize long-duration calculations. This study would not be possible without astronomical data from the VizieR (CDS, Strasbourg, France), SDSS (Sloan Digital Sky Survey), STScI MAST (Mikulski Archive for Space Telescopes), and IRSA (Infrared Science Archive) databases. We are grateful to the developers of these resources. In conclusion, we wish to thank the referee for a careful reading of the paper, helpful criticism, and a discussion of the results.


\newpage
\begin{thebibliography}{99}

\bibitem[Alam et al. (2015)]{Alam2015}
Alam et al. (Sh. Alam, F. D. Albareti, C. Allende Prieto, et al.), eprint arXiv: \textbf{1501.00963},  (2015).

\bibitem[van Altena et al. (1995)]{Altena1995}
van Altena et al. (W. F. van Altena, J. T. Lee, and E. D. Hoffleit), \textit{The General Catalogue of Trigonometric Stellar Parallaxes} (Yale Univ. Observ., New Haven, 1995).

\bibitem[Bergeron et al. (2011)]{Bergeron2011}
Bergeron et al. (P. Bergeron, F. Wesemael, P. Dufour, A. Beauchamp, C. Hunter, R. A. Saffer, A. Gianninas, M. T. Ruiz, et al.), Astrophys. J. \textbf{737}, 28 (2011).

\bibitem[Bressan et al. (2012)]{Bressan2012}
Bressan et al. (A. Bressan, P. Marigo, L. Girardi, B. Salasnich,
C. dal Cero, S. Rubele, and A. Nanni), Mon. Not. R. Astron. Soc. \textbf{427}, 127 (2012).

\bibitem[Chen et al. (2014)]{Chen2014}
Chen et al. (Y. Chen, L. Girardi, A. Bressan, P. Marigo, M. Barbieri, and X. Kong), Mon. Not. R. Astron. Soc. \textbf{444}, 2525 (2014).

\bibitem[Cutri et al. (2003)]{Cutri2003}
Cutri et al. (R. M. Cutri, M. F. Skrutskie, S. van Dyk, C. A. Beichman, J. M. Carpenter, T. Chester, L. Cambresy, T. Evans, et al.), \textit{The IRSA 2MASS All-Sky Point Source Catalog}, NASA/IPAC Infrared Science Archive (2003). http://irsa.ipac.caltech.edu/applications/Gator/

\bibitem[Dittmann et al. (2014)]{Dittmann2014}
Dittmann et al. (J. A. Dittmann, J. M. Irwin, D. Charbonneau, and Z. K. Berta-Thompson), Astrophys. J. \textbf{784}, 156 (2014).

\bibitem[Dunn et al. (1955)]{Dunn1995}
Dunn et al. (R. B. Dunn, H. Hugues, and W. J. Luyten), Astron. J. \textbf{60}, 274 (1955).

\bibitem[de Gennaro et al. (2008)]{Gennaro2008}
de Gennaro et al. (S. de Gennaro, T. von Hippel, D. E. Winget, S. O. Kepler, A. Nitta, D. Koester, and L. Althaus), Astron. J. \textbf{135}, 1 (2008).

\bibitem[Grosheva (2006)]{Grosheva2006}
Grosheva (E. A. Grosheva), Astrophysics \textbf{49}, 397 (2006).

\bibitem[Holberg and Bergeron (2006)]{Holberg2006}
Holberg and Bergeron (J. B. Holberg and P. Bergeron), Astrophys. J. \textbf{132}, 1221 (2006).

\bibitem[Izmailov et al. (2010)]{Izmailov2010}
Izmailov et al. (I. S. Izmailov, M. L. Khovricheva, M. Yu. Khovrichev, O. V. Kiyaeva, E. V. Khrutskaya, L. G. Romanenko, E. A. Grosheva, K. L. Maslennikov, and O. A. Kalinichenko), Astron. Lett. \textbf{36}, 349 (2010).

\bibitem[Khovritchev et al. (2013)]{Khovritchev2013}
Khovritchev et al. (M. Yu. Khovritchev, I. S. Izmailov, and E. V. Khrutskaya), Mon. Not. R. Astron. Soc. \textbf{435}, 1083 (2013).

\bibitem[Khrutskaya et al. (2011)]{Khrutskaya2011}
Khrutskaya et al. (E. V. Khrutskaya, A. A. Berezhnoi, and M. Yu. Khovrichev), Astron. Lett. \textbf{37}, 420 (2011).

\bibitem[Kowalski and Saumon (2006)]{Kowalski2006}
Kowalski and Saumon (P. M. Kowalski and D. Saumon), Astrophys. J. \textbf{651}, L137 (2006).

\bibitem[Lasker et al. (1998)]{Lasker1998}
Lasker et al. (B. M. Lasker, G. R. Greene, M. J. Lattanzi, et al.), \textit{Astrophysics and Algorithms: A DIMACS Workshop on Massive Astronomical Data Sets} (1998).

\bibitem[L{\'e}pine and Shara (2005)]{Lepine2005}
L{\'e}pine and Shara (S. L{\'e}pine and M. M. Shara), Astron. J. \textbf{129}, 1483 (2005).

\bibitem[Luyten (1979)]{Luyten1979}
Luyten (W. J. Luyten), \textit{New Luyten Catalogue of Stars with Proper Motions Larger than Two Tenths of an Arcsecond}, NLTT, with 1st Suppl. (University of Minnesota, Minneapolis, 1979).

\bibitem[Mason et al. (2001)]{Mason2001}
Mason et al. (B. D. Mason, G. L. Wycoff, W. I. Hartkopf, G. G. Douglass, and C. E. Worley), Astron. J. \textbf{122}, 3466 (2001).

\bibitem[Massey and Refregier (2005)]{Massey2005}
Massey et al. ((R. Massey and A. Refregier), Mon. Not. R. Astron. Soc. \textbf{363}, 197 (2005).

\bibitem[Michalik et al. (2014)]{Michalik2014}
Michalik et al. (D. Michalik, L. Lindegren, and D. Hobbs), Astron. Astrophys. \textbf{571}, A85 (2014).

\bibitem[Ochsenbein et al. (2000)]{Ochsenbein2000}
Ochsenbein et al. (F. Ochsenbein, P. Bauer, and J. Marcout), Astron. Astrophys. Suppl. Ser. \textbf{143}, 23 (2000).

\bibitem[Riedel et al. (2014)]{Riedel2014}
Riedel et al. (A. R. Riedel, Ch. T. Finch, T. J. Henry, J. P. Subasavage, W.-C. Jao, L. Malo, D. R. Rodriguez, R. J. White, et al.), Astron. J. \textbf{147}, 85 (2014).

\bibitem[Robin et al. (2003)]{Robin2003}
Robin et al. (A. C. Robin, C. Reyl, S. Derri’re, and S. Picaud), Astron. Astrophys. \textbf{409}, 523 (2003).

\bibitem[Sanders and Binney (2015)]{Sanders2015}
Sanders and Binney (J. Sanders and J. Binney), Mon. Not. R. Astron. Soc. \textbf{449}, 3479 (2015).

\bibitem[S{\"o}derhjelm (2005)]{Soderhjelm2005}
S{\"o}derhjelm (S. S{\"o}derhjelm ), ESA SP-576 (2005).

\bibitem[Spada et al. (2013)]{Spada2013}
Spada et al. (F. Spada, P. Demarque, Y.-C. Kim, and A. Sills), Astrophys. J. \textbf{776}, 87 (2013).

\bibitem[Tang et al. (2015)]{Tang2015}
Tang et al. (J. Tang, A. Bressan, P. Rosenfield, et al.), Mon. Not. R. Astron. Soc. \textbf{445}, 4287 (2015).

\bibitem[Thies et al. (2015)]{Thies2015}
Thies et al. (I. Thies, J. Pflamm-Altenburg, P. Kroupa, and M. Marks), Astrophys. J. \textbf{800}, 72 (2015).

\bibitem[Tremblay et al. (2011)]{Tremblay2011}
Tremblay et al. (P.-E. Tremblay, P. Bergeron, and A. Gianninas), Astrophys. J. \textbf{730}, 128 (2011).

\bibitem[Wielen et al. (1999)]{Wielen1999}
Wielen et al. (R. Wielen, C. Dettbarn, H. Jahreiss, H. Lenhardt, and H. Schwan), Astron. Astrophys. \textbf{346}, 675 (1999).

\bibitem[Wright et al. (2010)]{Wright2010}
Wright et al. (E. L. Wright, P. R. M. Eisenhardt, A. K. Mainzer, M. E. Ressler, R. M. Cutri, T. Jarrett, J. D. Kirkpatrick, D. Padgett, et al.), Astrophys. J. \textbf{140}, 1868 (2010).

\bibitem[Zacharias et al. (2013)]{Zacharias2013}
Zacharias et al. (N. Zacharias, C. T. Finch, T. M. Girard, A. Henden, J. L. Bartlett, D. G. Monet, and M. I. Zacharias), Astron. J. \textbf{145}, 44 (2013).

\bibitem[Zenoviene et al. (2015)]{Zenoviene2015}
Zenoviene et al. (R. Zenoviene, G. Tautvaisiene, B. Nordstr{\"o}m, E. Stonkut’, and G. Barisevi’ius), Astron. Astrophys. \textbf{576}, A113 (2015).

\end{thebibliography}
\end{document}